\documentclass{pasj00}
\draft

\usepackage{multirow}

\begin{document}
\SetRunningHead{Onozato \etal }{MIR sources that dramatically brightened}

\title{A Study of Mid-Infrared Sources that dramatically brightened}

\author{Hiroki \textsc{Onozato} and Yoshifusa \textsc{Ita}}
\affil{Astronomical Institute, Graduate School of Science, Tohoku University, 6-3 Aramaki Aoba, Aoba-ku, Sendai, Miyagi 980-0857}
\email{h.onozato@astr.tohoku.ac.jp}
\author{Kenji \textsc{Ono}}
\affil{Institute for Cosmic Ray Research, The University of Tokyo, 5-1-5 Kashiwa-no-Ha, Kashiwa, Chiba 277-8582}
\author{Misato \textsc{Fukagawa}}
\affil{Department of Earth and Space Science, Graduate School of Science Osaka University, 1-1 Machikaneyama, Toyonaka, Osaka 560-0043}
\author{Kenshi \textsc{Yanagisawa} and Hideyuki \textsc{Izumiura}}
\affil{Okayama Astrophysical Observatory, National Astronomical Observatory of Japan, 3037-5 Honjo, Kamogata, Asakuchi, Okayama 719-0232}
\author{Yoshikazu \textsc{Nakada}}
\affil{Kiso Observatory, Institute of Astronomy, School of Science, The University of Tokyo, 10762-30 Mitake, Kiso-machi, Kiso-gun, Nagano 397-0101}
\and
\author{Noriyuki \textsc{Matsunaga}}
\affil{Department of Astronomy, School of Science, The University of Tokyo, 7-3-1 Hongo, Bunkyo-ku, Tokyo 113-0033}


\maketitle

\begin{abstract}
We present results of near-infrared photometric and spectroscopic observations of mid-infrared (MIR) sources that dramatically brightened. Using IRAS, AKARI, and WISE point source catalogs, we found that 4 sources (IRAS~19574+491, V2494~Cyg, IRAS~22343+7501, and V583~Cas) significantly brightened at MIR wavelengths over the 20-30 years of difference in observing times. Little is known about these sources except V2494~Cyg, which is considered a FU Orionis star. Our observation clearly resolves IRAS~22343+7501 into 4 stars (2MASS~J22352345+7517076, 2MASS~J22352442+7517037, \textrm{[}RD95\textrm{]}~C, and 2MASS~J22352497+7517113) and first JHK$\mathrm{_{s}}$ photometric data for all 4 sources are obtained. Two of these stars (2MASS~J22352442+7517037 and 2MASS~J22352497+7517113) are known as T Tau stars. Our spectroscopic observation reveals that IRAS~19574+9441 is an M-type evolved star and V583~Cas is a carbon star. 2MASS~J22352345+7517076 is probably a YSO, judging from our observation that it has featureless near-infrared (NIR) spectrum and also showed dramatic brightening in NIR (about 4 magnitudes in K$\mathrm{_{s}}$-band). The possible reasons for dramatic brightening in MIR are discussed in this paper.
\end{abstract}

\section{Introduction}

Sensitive observations at mid-infrared wavelength regions have been conducted with satellites. Infrared Astronomical Satellite (IRAS, \cite{N}), AKARI \citep{Is}, and Wide-field Infrared Survey Explorer (WISE, \cite{Wr}) were launched in 1983, 2006, and 2009, respectively, and performed all-sky surveys. Infrared Space Observatory \citep{Ke}, Midcourse Space Experiment \citep{Pr}, and Spitzer Space Telescope were launched in 1995, 1996, and 2003, respectively, and observed many objects. Although less sensitive, there are ground-based mid-infrared (MIR) data.

Using these data, some astronomers have discussed MIR light variations. \citet{Mei} conducted Surveying the Agents of a Galaxy's Evolution (SAGE) of the Large Magellanic Cloud with Spitzer Space Telescope. \citet{V} detected many near-infrared (NIR) and MIR variable objects using the SAGE data taken over 3 months of difference in observing times. In the data, a large fraction of the extreme asymptotic giant branch (AGB) stars showed light variation. They also detected variable young stellar object (YSO) candidates. \citet{Mel} reported dramatic MIR flux decrease in young, Sun-like star (TYC 8241 2652 1), using AKARI and WISE data and by their own MIR observations with Thermal-Region Camera Spectrograph at the Gemini South telescope. They suggested that the reason for the large light variation is a disappearance of a warm, dusty circumstellar disk, but how the disk disappeared is currently not known. \citet{Ga} combined IRAS, Two Micron All-Sky Survey (2MASS) and WISE point source catalogs (PSCs) and found an infrared variable source. Based on the photometric characteristics of the object, they concluded that the object is likely to be a Sakurai's object \citep{Du}. \citet{Reb} summarized previous studies of NIR and MIR variation in young stars and \citet{Ho} reviewed the researches of NIR and MIR variation in cataclysmic variables with Spitzer Space Telescope data.

As described above, some papers that discussed MIR light variations have been published. However, the number of papers discussing MIR light variation is much smaller than that of papers on optical or NIR light variations because accurate MIR observation is rather difficult. We need to observe MIR or far-infrared (FIR) light variation for studying variations in circumstellar dust because the peak of dust emission is in MIR or FIR region. Additionally, most of the previous works studied short-term light variations (mainly using Spitzer Space Telescope) and selected targets by setting some selection criteria. For example, \citet{Ga} selected targets by 2MASS and WISE color.

\begin{figure}
  \begin{center}
    \FigureFile(80mm,48mm){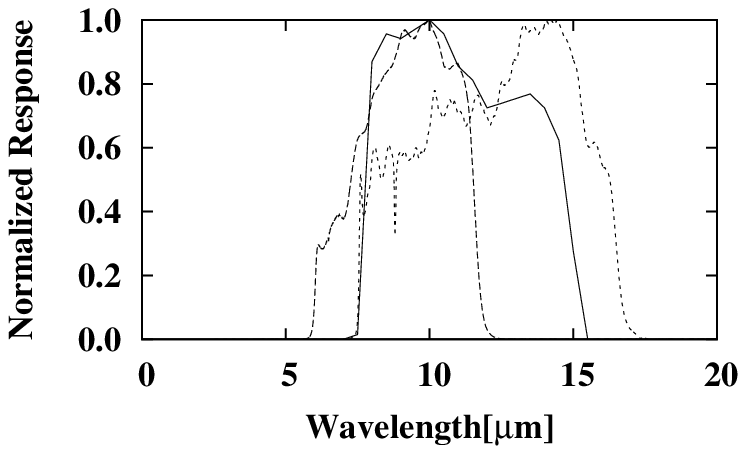}
    \FigureFile(80mm,48mm){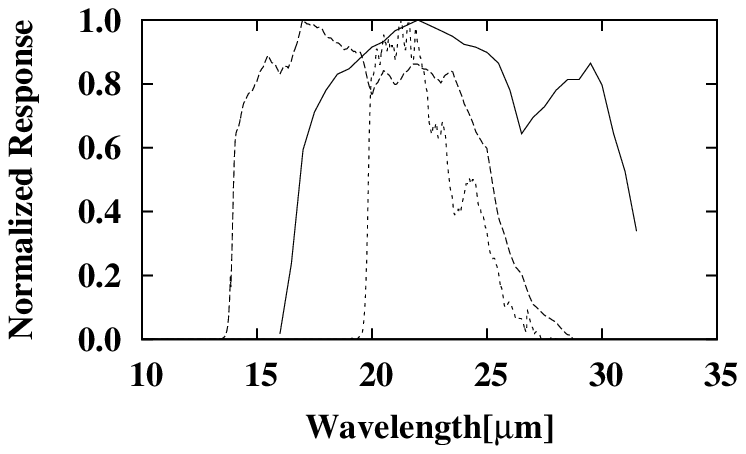}
  \end{center}
  \caption{Normalized response curve. Left: The solid curve corresponds to IRAS 12, the dashed curve to AKARI S9W, and the dotted curve to WISE W3. Right: The solid line represents IRAS 25, the dashed curve AKARI L18W, and the dotted curve WISE W4.}
  \label{fig:RC}
\end{figure}

In this paper, we take advantage of the fact that IRAS, AKARI and WISE performed all-sky surveys at similar wavelengths (IRAS 12, AKARI S9W, and WISE W3 and IRAS 25, AKARI L18W, and WISE W4, see figure \ref{fig:RC}). We combined these PSCs and searched the objects whose AKARI or WISE MIR fluxes had significantly increased from that of IRAS over the 20-30 years of difference in observing times. The aim of this study is to search sources that show significant MIR brightenings, and to know what types of objects show MIR large light variations and also to know why they show such large light variations.

\section{Data}

\subsection{Target Selection}

At first, we merged AKARI/IRC All-Sky Survey Point Source Catalog (Version 1.0, \cite{Is}) and WISE All-Sky Data Release \citep{Cu2012} with a tolerance angular radius of 5 arcsec. If more than one WISE source were found, we regarded the closest one as the identical source. We call it AKARI-WISE catalog. Combining AKARI and WISE PSCs is necessary to ensure that the sources are securely detected by both AKARI and WISE satellites. Then, we cross-correlated IRAS PSC, Version 2.0 \citep{Hel} with the AKARI-WISE catalog with a tolerance angular radius of 5 arcsec. The beam size of IRAS is about 10 times larger than those of AKARI and WISE, making the choice of the tolerance radius difficult. If we apply a small radius, the number of the successful cross identification will be small. If we use a large radius, we will suffer from miss-identifications. We took 5 arcsec as a compromise size between these two problems. When we applied 5 arcsec as a tolerance radius, we did not find the source that had another sources within the tolerance radius. Using the IRAS-AKARI-WISE combined catalog, we chose our sample stars by following criteria.
\begin{enumerate}
  \item The flux density quality of IRAS at 12 $\micron$ or 25 $\micron$ should be 3
  \item The signal-to-noise ratios (S/N) of IRAS, AKARI and WISE data should be better than 3
  \item Declination of the sources are larger than \timeform{-30D} so that they can be observed from northern hemisphere
\end{enumerate}
The reason we set the criteria of the flux density quality of IRAS and the S/N of IRAS, AKARI and WISE data is to discuss sources with significant flux increasing. The number of sources at each step of choosing our sample is shown in table \ref{tab:NS}.

\begin{figure}
 \begin{center}
  \FigureFile(80mm,48mm){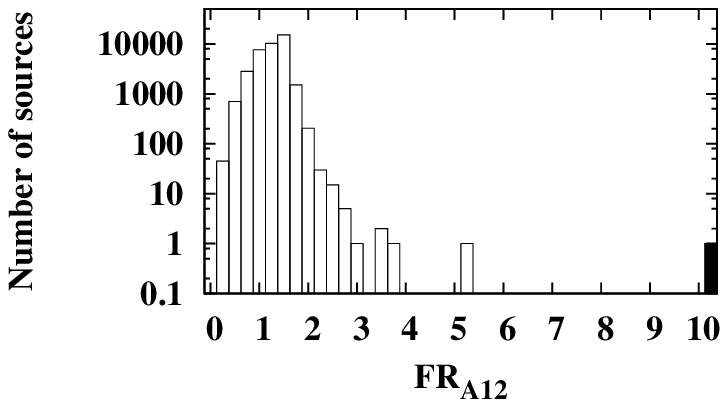}
  \FigureFile(80mm,48mm){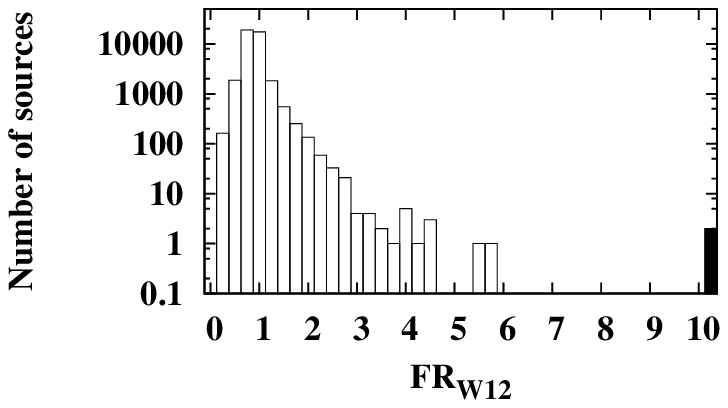}
  \FigureFile(80mm,48mm){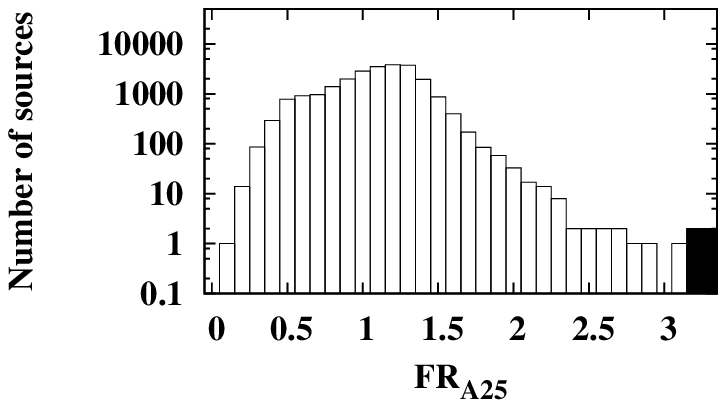}
  \FigureFile(80mm,48mm){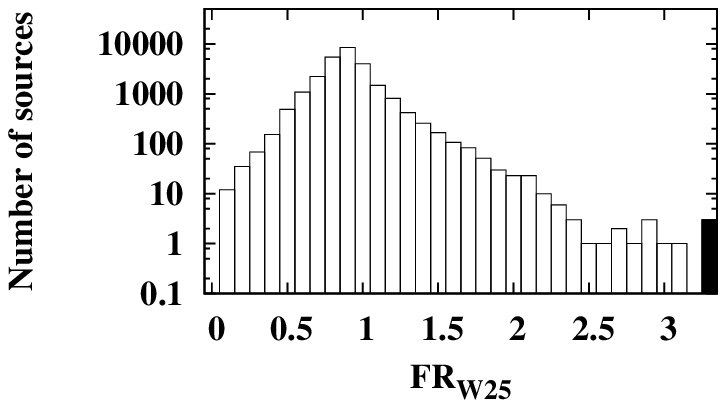}
 \end{center}
 \caption{The histograms of flux density ratios of each band. The empty histograms represent all samples selected by 4 steps and the black histograms show our 4 targets (for each step and the number of sources, see table \ref{tab:NS}). The rightmost histograms of each figure include all the sources whose $FR_{\mathrm{A12}}$ or $FR_{\mathrm{W12}}\ >\ 10.0$, or $FR_{\mathrm{A25}}$ or $FR_{\mathrm{W25}}\ > 3.2$. As can be seen in this figure, objects selected are very rare.}
 \label{fig:FRH}
\end{figure}

The flux density ratios between IRAS and AKARI or WISE versus the numbers of sources with these flux density ratio are shown in figure \ref{fig:FRH}, where
\begin{eqnarray}
  FR_{\mathrm{A12}}=\frac{F_{\nu,\: \mathrm{AKARI\ S9W}}}{F_{\nu,\: \mathrm{IRAS\ 12}}}\\
  FR_{\mathrm{W12}}=\frac{F_{\nu,\: \mathrm{WISE\ W3}}}{F_{\nu,\: \mathrm{IRAS\ 12}}}\\
  FR_{\mathrm{A25}}=\frac{F_{\nu,\: \mathrm{AKARI\ L18W}}}{F_{\nu,\: \mathrm{IRAS\ 25}}}\\
  FR_{\mathrm{W25}}=\frac{F_{\nu,\: \mathrm{WISE\ W4}}}{F_{\nu,\: \mathrm{IRAS\ 25}}}
\end{eqnarray}
and $F_{\nu}$ means flux density. To select only the sources with strongest brightening, we further applied the following criteria.
\begin{enumerate}
  \item $FR_{\mathrm{A12}}\ >$ 10
  \item $FR_{\mathrm{W12}}\ >$ 10
  \item $FR_{\mathrm{A25}}\ >\ \sqrt{10}\ \sim$ 3.16
  \item $FR_{\mathrm{W25}}\ >\ \sqrt{10}\ \sim$ 3.16
\end{enumerate}
These numbers were chosen rather arbitrary for this pilot study, so there is no physical meaning for the numbers. But as can be seen in figure \ref{fig:FRH}, the above criteria successfully chose sources showing the largest brightness changes. As a result, 4 sources were chosen. In the selection process, we did not apply color correction. 

In this paper, we do not discuss the sources whose AKARI or WISE's fluxes {\it decreased} from that of IRAS, because spatial resolution of IRAS is far worse than that of AKARI or WISE. Because of poor spatial resolution, it is possible that IRAS detected some point sources as one object which AKARI and WISE were able to resolve. In such cases, IRAS always overestimate fluxes compared to AKARI and WISE.

\begin{figure}
 \begin{center}
  \FigureFile(160mm,96mm){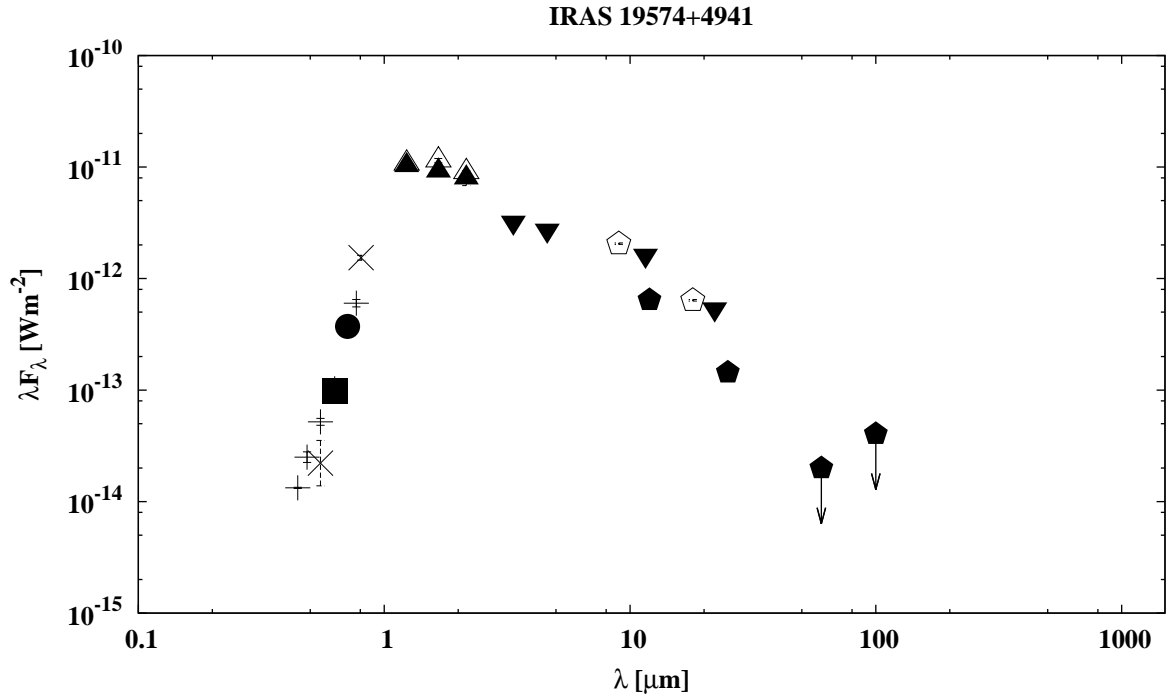}
 \end{center}
 \caption{The SED of IRAS~19574+4941. Meanings of the shapes of marks are given in table \ref{tab:PS}. Errors in observed fluxes are indicated, but it is hard to see because they are usually smaller than the size of marks. Points with arrows represent the data are upper limit.}
 \label{fig:SED1}
\end{figure}

\begin{figure}
 \begin{center}
  \FigureFile(160mm,96mm){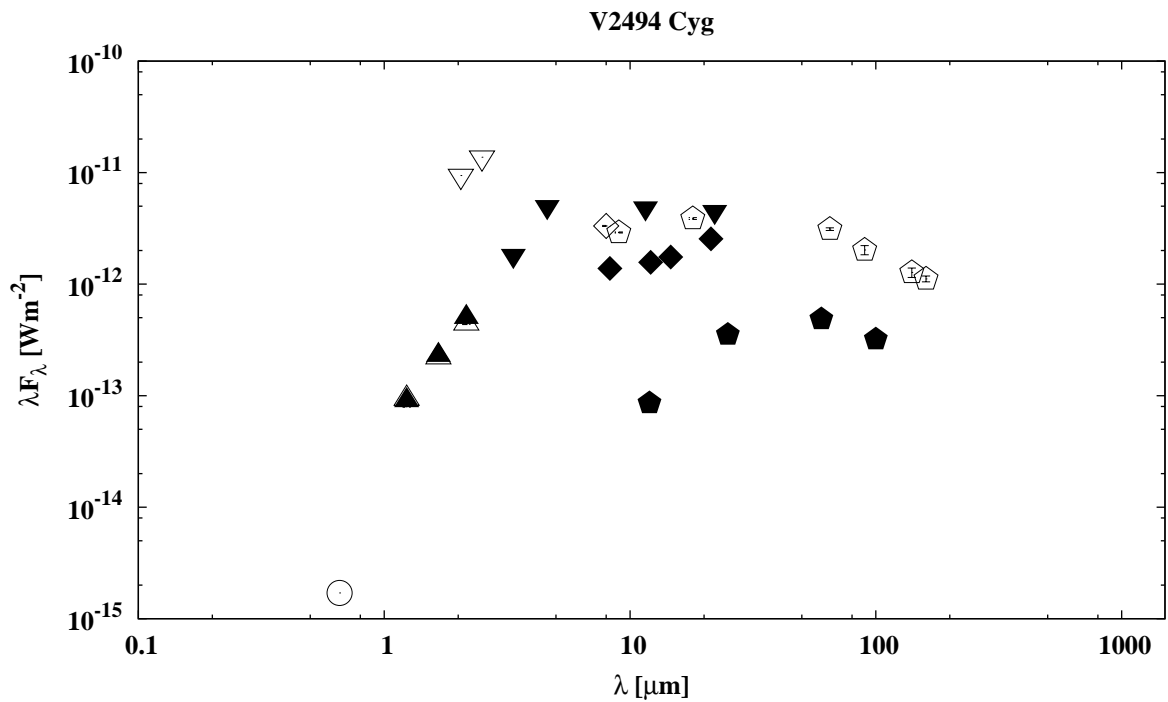}
 \end{center}
 \caption{Same as figure \ref{fig:SED1}, but for V2494~Cyg.}
 \label{fig:SED2}
\end{figure}

\begin{figure}
 \begin{center}
  \FigureFile(160mm,96mm){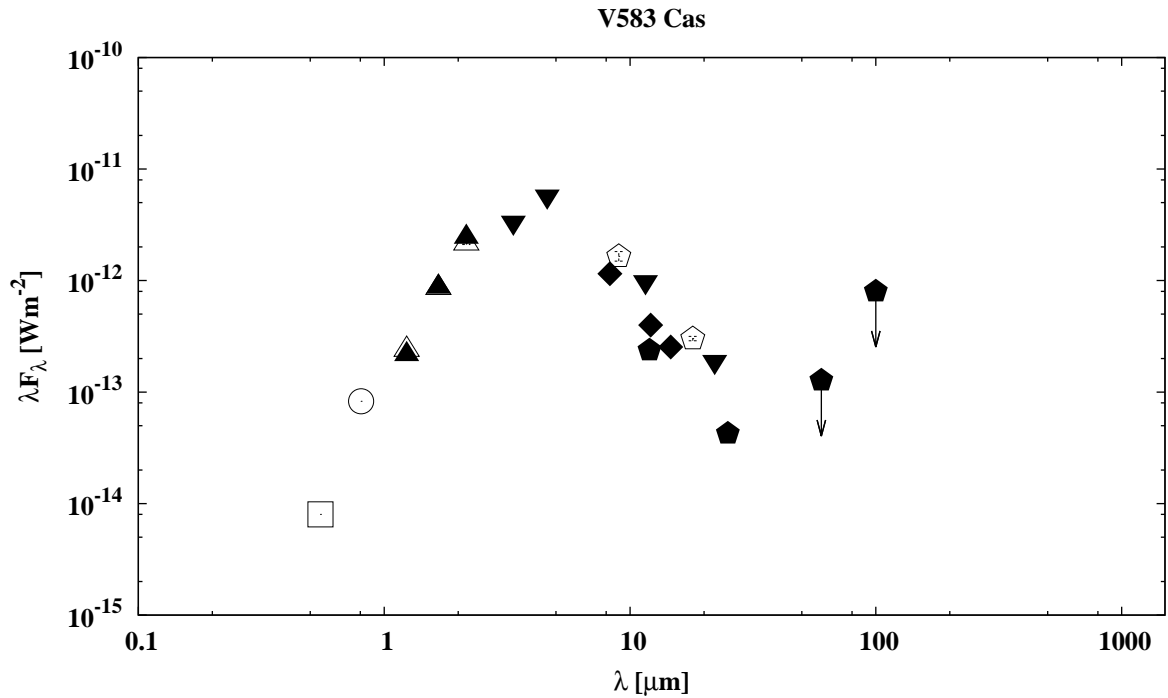}
 \end{center}
 \caption{Same as figure \ref{fig:SED1}, but for V583~Cas.}
 \label{fig:SED3}
\end{figure}

We chose 4 sources as our targets. The flux density ratios of these sources are shown in table \ref{tab:FR}. Basic information of them is shown in table \ref{tab:LOO}. No other sources are present within 5 arcsec around these 4 sources in IRAS catalog and AKARI-WISE catalog. However, as we show later, one source (IRAS~22343+7501) is resolved into 4 stars (2MASS~J22352345+7517076, 2MASS~J22352442+7517037, \textrm{[}RD95\textrm{]}~C, and 2MASS~J22352497+7517113) by \citet{Cu2003}, \citet{Kun}, and our observation. These sources are too close to be detected individually by IRAS, AKARI, and WISE. We made spectral energy distributions (SEDs) of our targets except for IRAS~22343+7501, using all available data to date and also data from our observation (see tables \ref{tab:IRAS19574}-\ref{tab:V583Cas}). They are shown in figures \ref{fig:SED1}-\ref{fig:SED3}. Meanings of the shapes of marks are given in table \ref{tab:PS}. Errors in observed fluxes are indicated, but it is hard to see these errors because they are smaller than the size of marks. Here, we stress that MIR brightening is significant.

\begin{table}
  \caption{The number of sources at each step of choosing our sample.}\label{tab:NS}
  \begin{center}
    \scalebox{0.90}
    {
      \begin{tabular*}{189mm}{lcccc}
        \hline
        \multicolumn{1}{c}{Step} & \multicolumn{1}{c}{AKARI S9W} & \multicolumn{1}{c}{WISE W3} & \multicolumn{1}{c}{AKARI L18W} & \multicolumn{1}{c}{WISE W4}\\
        \hline
        1. Merging AKARI and WISE catalogs & 839438 & 863972 & 191747 & 864130 \\
        2. Merging IRAS and AKARI-WISE catalogs & 57973 & 61944 & 54401 & 61938\\
        3. Choosing sources whose declinations are larger than \timeform{-30D} & 33926 & 36545 & 31267 & 36547\\
        4. Choosing sources whose S/N are better than 3, & \multirow{2}{*}{32677} & \multirow{2}{*}{35084} & \multirow{2}{*}{20588} & \multirow{2}{*}{22054}\\
        \ \ \ \ and IRAS flux density qualities are 3 & & & & \\
        \hline
      \end{tabular*}
    }
  \end{center}
\end{table}

\begin{table}
  \caption{Flux density ratios of selected sources.}\label{tab:FR}
  \begin{center}
    \begin{tabular}{lcccc}
      \hline
      \multicolumn{1}{c}{Target} & \multicolumn{1}{c}{$FR_{\mathrm{A12}}$} & \multicolumn{1}{c}{$FR_{\mathrm{W12}}$} & \multicolumn{1}{c}{$FR_{\mathrm{A25}}$} & \multicolumn{1}{c}{$FR_{\mathrm{W25}}$}\\
      \hline
      IRAS~19574+4941 & 2.4 $\pm$ 0.1 & 2.4 $\pm$ 0.1 & 3.2 $\pm$ 0.2 & 3.3 $\pm$ 0.2 \\
      V2494~Cyg & 25.4 $\pm$ 1.6 & 54.5 $\pm$ 3.3 & 7.9 $\pm$ 0.4 & 11.3 $\pm$ 0.6\\
      IRAS~22343+7501 & 3.3 $\pm$ 0.6 & 12.1 $\pm$ 0.5 & 1.3 $\pm$ 0.2 & 3.1 $\pm$ 0.2\\
      V583~Cas & 5.2 $\pm$ 0.6 & 4.0 $\pm$ 0.3 & 5.1 $\pm$ 0.5 & 3.9 $\pm$ 0.3\\
      \hline
    \end{tabular}
  \end{center}
\end{table}


\begin{table}
  \caption{List of observed stars.}\label{tab:LOO}
  \begin{center}
    \scalebox{0.61}
      {
      \begin{tabular*}{278mm}{lcccccl}
        \hline
        \multicolumn{1}{c}{Target names} & \multicolumn{1}{c}{Right Ascension\footnotemark[$*$]} & \multicolumn{1}{c}{Declination\footnotemark[$*$]} & \multicolumn{1}{c}{IRAS} & \multicolumn{1}{c}{AKARI} & \multicolumn{1}{c}{WISE} & \multicolumn{1}{c}{Other names}\\
        & \multicolumn{1}{c}{(J2000)} & \multicolumn{1}{c}{(J2000)} & & & & \\
        \hline
        IRAS~19574+4941 & 19 58 54.895 & +49 49 49.67 & 19574+4941 & 1958549+494949 & 195854.89+494949.4 &\\
        V2494~Cyg & 20 58 21.093 & +52 29 27.70 & 20568+5217 & 2058211+522927 & 205821.09+522927.6 & HH 381 IRS, MSX6C G091.6383+04.4007, WB89 19\\
        2MASS~J22352345+7517076 & 22 35 23.458 & +75 17 07.63 & 22343+7501 & 2235235+751708 & 223523.61+751708.3 & SSTc2d J223523.4+751708\\
        2MASS~J22352442+7517037 & 22 35 24.425 & +75 17 03.73 & 22343+7501 & 2235235+751708 & 223523.61+751708.3 & e.g. \textrm{[}KP93\textrm{]} 3-8, \textrm{[}RD95\textrm{]}~B, SSTc2d J223524.4+751704\\
        \textrm{[}RD95\textrm{]}~C & & & 22343+7501 & 2235235+751708 & 223523.61+751708.3 & \\
        2MASS~J22352497+7517113 & 22 35 24.978 & +75 17 11.37 & 22343+7501 & 2235235+751708 & 223523.61+751708.3 & e.g. \textrm{[}RD95\textrm{]}~A, \textrm{[}S2006\textrm{]} 8, XMMU J223524.6+751710\\
        V583~Cas & 23 31 16.203 & +61 32 17.96 & 23289+6115 & 2331162+613218 & 233116.24+613218.3 & CGCS 5887, NIKC 4-136\\
        \hline
        \multicolumn{7}{@{}l@{}}{\hbox to 0pt{\parbox{278mm}{\footnotesize
          Reference.
          \par\noindent
          \footnotemark[$*$] \citet{Cu2003}
        }\hss}}
      \end{tabular*}
    }
  \end{center}
\end{table}

\begin{table}
  \caption{Photometric data of IRAS~19574+4941.}\label{tab:IRAS19574}
  \begin{center}
    \begin{tabular}{cccccc}
      \hline
      \multicolumn{1}{c}{Band} & \multicolumn{1}{c}{Wavelength} & \multicolumn{1}{c}{$\lambda F_{\lambda}$} & \multicolumn{1}{c}{Epoch} & \multicolumn{1}{c}{Reference\footnotemark[$*$]}\\
      & \multicolumn{1}{c}{($\micron$)} & \multicolumn{1}{c}{$\mathrm{(Wm^{-2})}$} & & \\
      \hline
      B & 0.44 & $1.327\times 10^{-14}\pm 1.7\times 10^{-16}$ & & (1)\\
      g' & 0.49 & $2.503\times 10^{-14\ \ +2.92\times 10^{-15}}_{\ \ \ \ \ \ -2.61\times 10^{-15}}$ & & (1)\\
      V & 0.55 & $2.215\times 10^{-14\ \ +1.324\times 10^{-14}}_{\ \ \ \ \ \ -8.29\times 10^{-15}}$ & & (2)\\
      V & 0.55 & $5.186\times 10^{-14\ \ +3.81\times 10^{-15}}_{\ \ \ \ \ \ -3.55\times 10^{-15}}$ & & (1)\\
      r' & 0.63 & $9.801\times 10^{-14\ \ +5.76\times 10^{-15}}_{\ \ \ \ \ \ -3.55\times 10^{-15}}$ & 2001 August 11 & (5)\\
      r' & 0.63 & $1.026\times 10^{-13\ \ +6.7\times 10^{-15}}_{\ \ \ \ \ \ -6.3\times 10^{-15}}$ & & (1)\\
      R & 0.71 & $3.723\times 10^{-13}$ & & (7)\\
      i' & 0.77 & $6.015\times 10^{-13\ \ +4.84\times 10^{-14}}_{\ \ \ \ \ \ -4.48\times 10^{-15}}$ & & (1)\\
      $\mathrm{I_{C}}$ & 0.81 & $1.539\times 10^{-12\ \ +8.0\times 10^{-14}}_{\ \ \ \ \ \ -7.6\times 10^{-14}}$ & & (2)\\
      J & 1.2 & $1.069\times 10^{-11\ \ +2.5\times 10^{-13}}_{\ \ \ \ \ \ -2.4\times 10^{-13}}$ & 2000 May 30 & (8)\\
      J & 1.2 & $1.025\times 10^{-11\ \ +9.0\times 10^{-13}}_{\ \ \ \ \ \ -8.2\times 10^{-13}}$ & 2013 August 26 & (9)\\
      H & 1.7 & $1.141\times 10^{-11\ \ +5.0\times 10^{-13}}_{\ \ \ \ \ \ -4.8\times 10^{-13}}$ & 2000 May 30 & (8)\\
      H & 1.7 & $9.173\times 10^{-12\ \ +9.78\times 10^{-13}}_{\ \ \ \ \ \ -8.84\times 10^{-13}}$ & 2013 August 26 & (9)\\
      K$\mathrm{_{s}}$ & 2.2 & $8.964\times 10^{-12\ \ +2.17\times 10^{-13}}_{\ \ \ \ \ \ -2.12\times 10^{-13}}$ & 2000 May 30 & (8)\\
      K$\mathrm{_{s}}$ & 2.2 & $7.938\times 10^{-12\ \ +1.311\times 10^{-12}}_{\ \ \ \ \ \ -1.125\times 10^{-12}}$ & 2013 August 26 & (9)\\
      W1 & 3.4 & $3.216\times 10^{-12\ \ +2.11\times 10^{-13}}_{\ \ \ \ \ \ -1.98\times 10^{-13}}$ & 2010 & (11)\\
      W2 & 4.6 & $2.726\times 10^{-12\ \ +1.34\times 10^{-13}}_{\ \ \ \ \ \ -1.28\times 10^{-13}}$ & 2010 & (11)\\
      S9W & 9.0 & $2.007\times 10^{-12}\pm 1.1\times 10^{-14}$ & 2006-2007 & (15)\\
      W3 & 12 & $1.628\times 10^{-12\ \ +8\times 10^{-15}}_{\ \ \ \ \ \ -7\times 10^{-15}}$ & 2010 & (11)\\
      12 & 12 & $6.442\times 10^{-13}\pm 2.58\times 10^{-14}$ & 1983 & (17)\\
      L18W & 18 & $6.391\times 10^{-13}\pm 3.4\times 10^{-15}$ & 2006-2007 & (15)\\
      W4 & 22 & $5.373\times 10^{-13\ \ +4.5\times 10^{-15}}_{\ \ \ \ \ \ -4.4\times 10^{-15}}$ & 2010 & (11)\\
      25 & 25 & $1.443\times 10^{-13}\pm 8.7\times 10^{-15}$ & 1983 & (17)\\
      60 & 60 & $1.998\times 10^{-14}$ (Upper limit) & 1983 & (17)\\
      100 & 100 & $4.058\times 10^{-14}$ (Upper limit) & 1983 & (17)\\
      \hline
      \multicolumn{5}{@{}l@{}}{\hbox to 0pt{\parbox{170mm}{\footnotesize
        Notes.
        \par\noindent
        \footnotemark[$*$] Reference numbers are described in table \ref{tab:PS}.
      }\hss}}
    \end{tabular}
  \end{center}
\end{table}

\begin{table}
  \caption{Photometric data of V2494~Cyg.}\label{tab:V2494Cyg}
  \begin{center}
    \begin{tabular}{ccccc}
      \hline
      \multicolumn{1}{c}{Band} & \multicolumn{1}{c}{Wavelength} & \multicolumn{1}{c}{$\lambda F_{\lambda}$} & \multicolumn{1}{c}{Epoch} & \multicolumn{1}{c}{Reference\footnotemark[$*$]}\\
      & \multicolumn{1}{c}{($\micron$)} & \multicolumn{1}{c}{$\mathrm{(Wm^{-2})}$} & & \\
      \hline
      $\mathrm{R_{C}}$ & 0.66 & $1.700\times 10^{-15}$ & & (6)\\
      J & 1.2 & $9.333\times 10^{-14\ \ +2.70\times 10^{-15}}_{\ \ \ \ \ \ -2.63\times 10^{-15}}$ & 1999 June 21 & (8)\\
      J & 1.2 & $8.979\times 10^{-14\ \ +1.114\times 10^{-14}}_{\ \ \ \ \ \ -9.91\times 10^{-15}}$ & 2013 August 26 & (9)\\
      H & 1.7 & $2.194\times 10^{-13\ \ +6.1\times 10^{-15}}_{\ \ \ \ \ \ -6.0\times 10^{-15}}$ & 1999 June 21 & (8)\\
      H & 1.7 & $2.283\times 10^{-13\ \ +2.01\times 10^{-14}}_{\ \ \ \ \ \ -1.85\times 10^{-14}}$ & 2013 August 26 & (9)\\
      2.05 & 2.05 & $9.430\times 10^{-12}$ & 1996 November 17-19 & (10)\\
      K$\mathrm{_{s}}$ & 2.2 & $4.411\times 10^{-13\ \ +7.0\times 10^{-15}}_{\ \ \ \ \ \ -6.9\times 10^{-15}}$ & 1999 June 21 & (8)\\
      K$\mathrm{_{s}}$ & 2.2 & $5.004\times 10^{-13\ \ +6.42\times 10^{-14}}_{\ \ \ \ \ \ -5.69\times 10^{-14}}$ & 2013 August 26 & (9)\\
      2.50 & 2.50 & $1.375\times 10^{-11}$ & 1996 November 17-19 & (10)\\
      W1 & 3.4 & $1.817\times 10^{-12\ \ +1.07\times 10^{-13}}_{\ \ \ \ \ \ -1.01\times 10^{-13}}$ & 2010 & (11)\\
      W2 & 4.6 & $4.989\times 10^{-12\ \ +3.03\times 10^{-13}}_{\ \ \ \ \ \ -2.86\times 10^{-13}}$ & 2010 & (11)\\
      IRAC4 & 8.0 & $3.324\times 10^{-12}\pm 3.8\times 10^{-14}$ & 2008 June 29 & (13)\\
      A & 8.28 & $1.382\times 10^{-12}\pm 5.7\times 10^{-14}$ & 1996-1997 & (14)\\
      S9W & 9.0 & $2.908\times 10^{-12}\pm 3.1\times 10^{-14}$ & 2006-2007 & (15)\\
      W3 & 12 & $4.854\times 10^{-12}\pm 4.9\times 10^{-14}$ & 2010 & (11)\\
      12 & 12 & $8.587\times 10^{-14}\pm 6.55\times 10^{-15}$ & 1983 & (17)\\
      C & 12.13 & $1.571\times 10^{-12}\pm 8.3\times 10^{-14}$ & 1996-1997 & (14)\\
      D & 14.65 & $1.750\times 10^{-12}\pm 1.07\times 10^{-13}$ & 1996-1997 & (14)\\
      L18W & 18 & $3.879\times 10^{-12}\pm 7.3\times 10^{-14}$ & 2006-2007 & (15)\\
      E & 21.34 & $2.550\times 10^{-12}\pm 1.56\times 10^{-13}$ & 1996-1997 & (14)\\
      W4 & 22 & $4.506\times 10^{-12}\pm 2.9\times 10^{-14}$ & 2010 & (11)\\
      25 & 25 & $3.533\times 10^{-13}\pm 1.77\times 10^{-14}$ & 1983 & (17)\\
      60 & 60 & $4.873\times 10^{-13}\pm 3.41\times 10^{-14}$ & 1983 & (17)\\
      N60 & 65 & $3.113\times 10^{-12}\pm 8.6\times 10^{-14}$ & 2006-2007 & (16)\\
      WIDE-S & 90 & $2.023\times 10^{-12}\pm 1.91\times 10^{-13}$ & 2006-2007 & (16)\\
      100 & 100 & $3.208\times 10^{-13}\pm 5.10\times 10^{-14}$ & 1983 & (17)\\
      WIDE-L & 140 & $1.273\times 10^{-12}\pm 1.24\times 10^{-13}$ & 2006-2007 & (16)\\
      N160 & 160 & $1.116\times 10^{-12}\pm 6.7\times 10^{-14}$ & 2006-2007 & (16)\\
      \hline
      \multicolumn{5}{@{}l@{}}{\hbox to 0pt{\parbox{170mm}{\footnotesize
        Notes.
        \par\noindent
        \footnotemark[$*$] Reference numbers are described in table \ref{tab:PS}.
      }\hss}}
    \end{tabular}
  \end{center}
\end{table}

\begin{table}
  \caption{Photometric data of 2MASS~J22352345+7517076.}\label{tab:2MASSJ22352345}
  \begin{center}
    \begin{tabular}{ccccc}
      \hline
      \multicolumn{1}{c}{Band} & \multicolumn{1}{c}{Wavelength} & \multicolumn{1}{c}{$\lambda F_{\lambda}$} & \multicolumn{1}{c}{Epoch} & \multicolumn{1}{c}{Reference\footnotemark[$*$]}\\
      & \multicolumn{1}{c}{($\micron$)} & \multicolumn{1}{c}{$\mathrm{(Wm^{-2})}$} & & \\
      \hline
      J & 1.2 & $7.864\times 10^{-15}$ (Upper limit) & 1999 October 11 & (8)\\
      J & 1.2 & $3.675\times 10^{-15\ \ +8.42\times 10^{-16}}_{\ \ \ \ \ \ -6.85\times 10^{-16}}$ & 2013 August 26 & (9)\\
      H & 1.7 & $2.848\times 10^{-14}$ (Upper limit) & 1999 October 11 & (8)\\
      H & 1.7 & $1.019\times 10^{-13\ \ +1.38\times 10^{-14}}_{\ \ \ \ \ \ -1.22\times 10^{-14}}$ & 2013 August 26 & (9)\\
      K$\mathrm{_{s}}$ & 2.2 & $2.140\times 10^{-14\ \ +7.2\times 10^{-16}}_{\ \ \ \ \ \ -7.0\times 10^{-16}}$ & 1999 October 11 & (8)\\
      K$\mathrm{_{s}}$ & 2.2 & $8.281\times 10^{-13\ \ +1.019\times 10^{-13}}_{\ \ \ \ \ \ -9.07\times 10^{-14}}$ & 2013 August 26 & (9)\\
      W1 & 3.4 & $6.921\times 10^{-12\ \ +6.54\times 10^{-13}}_{\ \ \ \ \ \ -5.97\times 10^{-13}}$ & 2010 & (11)\\
      IRAC1 & 3.6 & $3.106\times 10^{-13}\pm 2.14\times 10^{-14}$ & 2004 April 4 & (12)\\
      IRAC2 & 4.5 & $6.662\times 10^{-13}\pm 7.59\times 10^{-14}$ & 2004 April 4 & (12)\\
      W2 & 4.6 & $2.312\times 10^{-11\ \ +5.6\times 10^{-13}}_{\ \ \ \ \ \ -5.5\times 10^{-13}}$ & 2010 & (11)\\
      IRAC3 & 5.8 & $9.614\times 10^{-13}\pm 6.15\times 10^{-14}$ & 2004 April 4 & (12)\\
      S9W & 9.0 & $5.495\times 10^{-12}\pm 1.039\times 10^{-12}$ & 2006-2007 & (15)\\
      W3 & 12 & $1.556\times 10^{-11\ \ +1.2\times 10^{-13}}_{\ \ \ \ \ \ -1.1\times 10^{-13}}$ & 2010 & (11)\\
      12 & 12 & $1.241\times 10^{-12}\pm 5.0\times 10^{-14}$ & 1983 & (17)\\
      L18W & 18 & $5.638\times 10^{-12}\pm 7.38\times 10^{-13}$ & 2006-2007 & (15)\\
      W4 & 22 & $1.086\times 10^{-11}\pm 1\times 10^{-14}$ & 2010 & (11)\\
      25 & 25 & $3.128\times 10^{-12}\pm 1.56\times 10^{-13}$ & 1983 & (17)\\
      60 & 60 & $3.316\times 10^{-12}\pm 2.98\times 10^{-13}$ & 1983 & (17)\\
      N60 & 65 & $3.706\times 10^{-12}\pm 4.77\times 10^{-13}$ & 2006-2007 & (16)\\
      MIPS2 & 72 & $1.977\times 10^{-12}\pm 4.37\times 10^{-13}$ & & (12)\\
      WIDE-S & 90 & $2.415\times 10^{-12}\pm 2.72\times 10^{-13}$ & 2006-2007 & (16)\\
      100 & 100 & $2.400\times 10^{-12}\pm 3.84\times 10^{-13}$ & 1983 & (17)\\
      WIDE-L & 140 & $2.069\times 10^{-12}\pm 1.74\times 10^{-13}$ & 2006-2007 & (16)\\
      N160 & 160 & $2.024\times 10^{-12}\pm 1.61\times 10^{-13}$ & 2006-2007 & (16)\\
      450 & 450 & $7.555\times 10^{-14}$ & & (18)\\
      800 & 800 & $2.661\times 10^{-15}\pm 4.27\times 10^{-16}$ & 1992 & (19)\\
      850 & 850 & $2.003\times 10^{-14}$ & & (18)\\
      1100 & 1100 & $1.044\times 10^{-15}\pm 8.2\times 10^{-17}$ & 1992 & (19)\\
      1300 & 1300 & $5.350\times 10^{-16}\pm 5.53\times 10^{-17}$ & 1992 & (19)\\
      \hline
      \multicolumn{5}{@{}l@{}}{\hbox to 0pt{\parbox{140mm}{\footnotesize
        Notes.
        \par\noindent
        \footnotemark[$*$] Reference numbers are described in table \ref{tab:PS}.
        \par\noindent
        \footnotemark[$\dagger$] 2MASS~J22352345+7517076, 2MASS~J22352442+7517037, \textrm{[}RD95\textrm{]}~C, and 2MASS~J22352497+7517113 were not separated without \citet{Kun}, 2MASS, ISLE, and IRAC (C2D Fall'07 High Reliability (HREL) CORES Catalog).
      }\hss}}
    \end{tabular}
  \end{center}
\end{table}

\begin{table}
  \caption{Photometric data of 2MASS~J22352442+7517037.}\label{tab:2MASSJ22352442}
  \begin{center}
    \begin{tabular}{ccccc}
      \hline
      \multicolumn{1}{c}{Band} & \multicolumn{1}{c}{Wavelength} & \multicolumn{1}{c}{$\lambda F_{\lambda}$} & \multicolumn{1}{c}{Epoch} & \multicolumn{1}{c}{Reference\footnotemark[$*$]}\\
      & \multicolumn{1}{c}{($\micron$)} & \multicolumn{1}{c}{$\mathrm{(Wm^{-2})}$} & & \\
      \hline
      V & 0.55 & $1.281\times 10^{-17}$ & 2004-2006 & (3)\\
      $\mathrm{R_{C}}$ & 0.66 & $1.155\times 10^{-16\ \ +2.99\times 10^{-17}}_{\ \ \ \ \ \ -2.37\times 10^{-17}}$ & 2004-2006 & (3)\\
      $\mathrm{I_{C}}$ & 0.81 & $7.140\times 10^{-16\ \ +2.272\times 10^{-16}}_{\ \ \ \ \ \ -1.724\times 10^{-16}}$ & 2004-2006 & (3)\\
      J & 1.2 & $2.540\times 10^{-14}$ (Upper limit) & 1999 October 11 & (8)\\
      J & 1.2 & $1.389\times 10^{-14\ \ +1.44\times 10^{-15}}_{\ \ \ \ \ \ -1.30\times 10^{-15}}$ & 2013 August 26 & (9)\\
      H & 1.7 & $6.744\times 10^{-14}$ (Upper limit) & 1999 October 11 & (8)\\
      H & 1.7 & $4.894\times 10^{-14\ \ +6.58\times 10^{-15}}_{\ \ \ \ \ \ -5.80\times 10^{-15}}$ & 2013 August 26 & (9)\\
      K$\mathrm{_{s}}$ & 2.2 & $1.026\times 10^{-13\ \ +3.5\times 10^{-15}}_{\ \ \ \ \ \ -3.3\times 10^{-15}}$ & 1999 October 11 & (8)\\
      K$\mathrm{_{s}}$ & 2.2 & $8.420\times 10^{-14\ \ +9.58\times 10^{-15}}_{\ \ \ \ \ \ -8.60\times 10^{-15}}$ & 2013 August 26 & (9)\\
      W1 & 3.4 & $6.921\times 10^{-12\ \ +6.54\times 10^{-13}}_{\ \ \ \ \ \ -5.97\times 10^{-13}}$ & 2010 & (11)\\
      W2 & 4.6 & $2.312\times 10^{-11\ \ +5.6\times 10^{-13}}_{\ \ \ \ \ \ -5.5\times 10^{-13}}$ & 2010 & (11)\\
      S9W & 9.0 & $5.495\times 10^{-12}\pm 1.039\times 10^{-12}$ & 2006-2007 & (15)\\
      W3 & 12 & $1.556\times 10^{-11\ \ +1.2\times 10^{-13}}_{\ \ \ \ \ \ -1.1\times 10^{-13}}$ & 2010 & (11)\\
      12 & 12 & $1.241\times 10^{-12}\pm 5.0\times 10^{-14}$ & 1983 & (17)\\
      L18W & 18 & $5.638\times 10^{-12}\pm 7.38\times 10^{-13}$ & 2006-2007 & (15)\\
      W4 & 22 & $1.086\times 10^{-11}\pm 1\times 10^{-14}$ & 2010 & (11)\\
      25 & 25 & $3.128\times 10^{-12}\pm 1.56\times 10^{-13}$ & 1983 & (17)\\
      60 & 60 & $3.316\times 10^{-12}\pm 2.98\times 10^{-13}$ & 1983 & (17)\\
      N60 & 65 & $3.706\times 10^{-12}\pm 4.77\times 10^{-13}$ & 2006-2007 & (16)\\
      MIPS2 & 72 & $1.977\times 10^{-12}\pm 4.37\times 10^{-13}$ & & (12)\\
      WIDE-S & 90 & $2.415\times 10^{-12}\pm 2.72\times 10^{-13}$ & 2006-2007 & (16)\\
      100 & 100 & $2.400\times 10^{-12}\pm 3.84\times 10^{-13}$ & 1983 & (17)\\
      WIDE-L & 140 & $2.069\times 10^{-12}\pm 1.74\times 10^{-13}$ & 2006-2007 & (16)\\
      N160 & 160 & $2.024\times 10^{-12}\pm 1.61\times 10^{-13}$ & 2006-2007 & (16)\\
      450 & 450 & $7.555\times 10^{-14}$ & & (18)\\
      800 & 800 & $2.661\times 10^{-15}\pm 4.27\times 10^{-16}$ & 1992 & (19)\\
      850 & 850 & $2.003\times 10^{-14}$ & & (18)\\
      1100 & 1100 & $1.044\times 10^{-15}\pm 8.2\times 10^{-17}$ & 1992 & (19)\\
      1300 & 1300 & $5.350\times 10^{-16}\pm 5.53\times 10^{-17}$ & 1992 & (19)\\
      \hline
      \multicolumn{5}{@{}l@{}}{\hbox to 0pt{\parbox{140mm}{\footnotesize
        Notes.
        \par\noindent
        \footnotemark[$*$] Reference numbers are described in table \ref{tab:PS}.
        \par\noindent
        \footnotemark[$\dagger$] 2MASS~J22352345+7517076, 2MASS~J22352442+7517037, \textrm{[}RD95\textrm{]}~C, and 2MASS~J22352497+7517113 were not separated without \citet{Kun}, 2MASS, ISLE, and IRAC (C2D Fall'07 High Reliability (HREL) CORES Catalog).
      }\hss}}
    \end{tabular}
  \end{center}
\end{table}

\begin{table}
  \caption{Photometric data of 2MASS~J22352497+7517113.}\label{tab:2MASSJ22352497}
  \begin{center}
    \begin{tabular}{ccccc}
      \hline
      \multicolumn{1}{c}{Band} & \multicolumn{1}{c}{Wavelength} & \multicolumn{1}{c}{$\lambda F_{\lambda}$} & \multicolumn{1}{c}{Epoch} & \multicolumn{1}{c}{Reference\footnotemark[$*$]}\\
      & \multicolumn{1}{c}{($\micron$)} & \multicolumn{1}{c}{$\mathrm{(Wm^{-2})}$} & & \\
      \hline
      V & 0.55 & $8.466\times 10^{-17}$ & 2004-2006 & (3)\\
      $\mathrm{R_{C}}$ & 0.66 & $3.180\times 10^{-16\ \ +3.224\times 10^{-16}}_{\ \ \ \ \ \ -1.601\times 10^{-16}}$ & 2004-2006 & (3)\\
      $\mathrm{I_{C}}$ & 0.81 & $3.324\times 10^{-15\ \ +2.453\times 10^{-15}}_{\ \ \ \ \ \ -1.411\times 10^{-16}}$ & 2004-2006 & (3)\\
      J & 1.2 & $9.333\times 10^{-14\ \ +2.70\times 10^{-15}}_{\ \ \ \ \ \ -2.63\times 10^{-15}}$ & 1999 October 11 & (8)\\
      J & 1.2 & $6.547\times 10^{-14\ \ +7.05\times 10^{-15}}_{\ \ \ \ \ \ -6.36\times 10^{-15}}$ & 2013 August 26 & (9)\\
      H & 1.7 & $2.194\times 10^{-13\ \ +6.0\times 10^{-15}}_{\ \ \ \ \ \ -6.0\times 10^{-15}}$ & 1999 October 11 & (8)\\
      H & 1.7 & $2.487\times 10^{-13\ \ +3.48\times 10^{-14}}_{\ \ \ \ \ \ -3.05\times 10^{-14}}$ & 2013 August 26 & (9)\\
      K$\mathrm{_{s}}$ & 2.2 & $4.411\times 10^{-13\ \ +7.0\times 10^{-15}}_{\ \ \ \ \ \ -6.9\times 10^{-15}}$ & 1999 October 11 & (8)\\
      K$\mathrm{_{s}}$ & 2.2 & $3.945\times 10^{-13\ \ +4.21\times 10^{-14}}_{\ \ \ \ \ \ -3.80\times 10^{-14}}$ & 2013 August 26 & (9)\\
      W1 & 3.4 & $6.921\times 10^{-12\ \ +6.54\times 10^{-13}}_{\ \ \ \ \ \ -5.97\times 10^{-13}}$ & 2010 & (11)\\
      W2 & 4.6 & $2.312\times 10^{-11\ \ +5.6\times 10^{-13}}_{\ \ \ \ \ \ -5.5\times 10^{-13}}$ & 2010 & (11)\\
      S9W & 9.0 & $5.495\times 10^{-12}\pm 1.039\times 10^{-12}$ & 2006-2007 & (15)\\
      W3 & 12 & $1.556\times 10^{-11\ \ +1.2\times 10^{-13}}_{\ \ \ \ \ \ -1.1\times 10^{-13}}$ & 2010 & (11)\\
      12 & 12 & $1.241\times 10^{-12}\pm 5.0\times 10^{-14}$ & 1983 & (17)\\
      L18W & 18 & $5.638\times 10^{-12}\pm 7.38\times 10^{-13}$ & 2006-2007 & (15)\\
      W4 & 22 & $1.086\times 10^{-11}\pm 1\times 10^{-14}$ & 2010 & (11)\\
      25 & 25 & $3.128\times 10^{-12}\pm 1.56\times 10^{-13}$ & 1983 & (17)\\
      60 & 60 & $3.316\times 10^{-12}\pm 2.98\times 10^{-13}$ & 1983 & (17)\\
      N60 & 65 & $3.706\times 10^{-12}\pm 4.77\times 10^{-13}$ & 2006-2007 & (16)\\
      MIPS2 & 72 & $1.977\times 10^{-12}\pm 4.37\times 10^{-13}$ & & (12)\\
      WIDE-S & 90 & $2.415\times 10^{-12}\pm 2.72\times 10^{-13}$ & 2006-2007 & (16)\\
      100 & 100 & $2.400\times 10^{-12}\pm 3.84\times 10^{-13}$ & 1983 & (17)\\
      WIDE-L & 140 & $2.069\times 10^{-12}\pm 1.74\times 10^{-13}$ & 2006-2007 & (16)\\
      N160 & 160 & $2.024\times 10^{-12}\pm 1.61\times 10^{-13}$ & 2006-2007 & (16)\\
      450 & 450 & $7.555\times 10^{-14}$ & & (18)\\
      800 & 800 & $2.661\times 10^{-15}\pm 4.27\times 10^{-16}$ & 1992 & (19)\\
      850 & 850 & $2.003\times 10^{-14}$ & & (18)\\
      1100 & 1100 & $1.044\times 10^{-15}\pm 8.2\times 10^{-17}$ & 1992 & (19)\\
      1300 & 1300 & $5.350\times 10^{-16}\pm 5.53\times 10^{-17}$ & 1992 & (19)\\
      \hline
      \multicolumn{5}{@{}l@{}}{\hbox to 0pt{\parbox{140mm}{\footnotesize
        Notes.
        \par\noindent
        \footnotemark[$*$] Reference numbers are described in table \ref{tab:PS}.
        \par\noindent
        \footnotemark[$\dagger$] 2MASS~J22352345+7517076, 2MASS~J22352442+7517037, \textrm{[}RD95\textrm{]}~C, and 2MASS~J22352497+7517113 were not separated without \citet{Kun}, 2MASS, ISLE, and IRAC (C2D Fall'07 High Reliability (HREL) CORES Catalog).
      }\hss}}
    \end{tabular}
  \end{center}
\end{table}

\begin{table}
  \caption{Photometric data of V583~Cas.}\label{tab:V583Cas}
  \begin{center}
    \begin{tabular}{ccccc}
      \hline
      \multicolumn{1}{c}{Band} & \multicolumn{1}{c}{Wavelength} & \multicolumn{1}{c}{$\lambda F_{\lambda}$} & \multicolumn{1}{c}{Epoch} & \multicolumn{1}{c}{Reference\footnotemark[$*$]}\\
     & \multicolumn{1}{c}{($\micron$)} & \multicolumn{1}{c}{$\mathrm{(Wm^{-2})}$} & & \\
     \hline
      V & 0.55 & $8.011\times 10^{-15}$ & & (4)\\
      $\mathrm{I_{C}}$ & 0.81 & $8.274\times 10^{-14}$ & & (6)\\
      J & 1.2 & $2.381\times 10^{-13\ \ +5.1\times 10^{-15}}_{\ \ \ \ \ \ -5.0\times 10^{-15}}$ & 1999 October 15 & (8)\\
      J & 1.2 & $2.158\times 10^{-13\ \ +8.9\times 10^{-15}}_{\ \ \ \ \ \ -8.6\times 10^{-15}}$ & 2013 August 26 & (9)\\
      H & 1.7 & $8.537\times 10^{-13\ \ +1.83\times 10^{-14}}_{\ \ \ \ \ \ -1.79\times 10^{-14}}$ & 1999 October 15 & (8)\\
      H & 1.7 & $8.679\times 10^{-13\ \ +4.17\times 10^{-14}}_{\ \ \ \ \ \ -3.98\times 10^{-14}}$ & 2013 August 26 & (9)\\
      K$\mathrm{_{s}}$ & 2.2 & $2.138\times 10^{-12\ \ +3.6\times 10^{-14}}_{\ \ \ \ \ \ -3.5\times 10^{-14}}$ & 1999 October 15 & (8)\\
      K$\mathrm{_{s}}$ & 2.2 & $2.435\times 10^{-12\ \ +1.77\times 10^{-13}}_{\ \ \ \ \ \ -1.65\times 10^{-13}}$ & 2013 August 26 & (9)\\
      W1 & 3.4 & $3.352\times 10^{-12\ \ +3.20\times 10^{-13}}_{\ \ \ \ \ \ -2.92\times 10^{-13}}$ & 2010 & (11)\\
      W2 & 4.6 & $5.754\times 10^{-12\ \ +6.31\times 10^{-13}}_{\ \ \ \ \ \ -5.69\times 10^{-13}}$ & 2010 & (11)\\
      A & 8.28 & $1.152\times 10^{-12}\pm 4.7\times 10^{-14}$ & 1996-1997 & (14)\\
      S9W & 9.0 & $1.655\times 10^{-12}\pm 1.62\times 10^{-13}$ & 2006-2007 & (15)\\
      W3 & 12 & $9.812\times 10^{-13\ \ +3.78\times 10^{-14}}_{\ \ \ \ \ \ -3.64\times 10^{-14}}$ & 2010 & (11)\\
      12 & 12 & $2.376\times 10^{-13}\pm 1.43\times 10^{-14}$ & 1983 & (17)\\
      C & 12.13 & $3.981\times 10^{-13}\pm 2.59\times 10^{-14}$ & 1996-1997 & (14)\\
      D & 14.65 & $2.540\times 10^{-13}\pm 1.82\times 10^{-14}$ & 1996-1997 & (14)\\
      L18W & 18 & $3.016\times 10^{-13}\pm 1.30\times 10^{-14}$ & 2006-2007 & (15)\\
      W4 & 22 & $1.894\times 10^{-13\ \ +4.9\times 10^{-15}}_{\ \ \ \ \ \ -4.8\times 10^{-15}}$ & 2010 & (11)\\
      25 & 25 & $4.255\times 10^{-14}\pm 3.40\times 10^{-15}$ & 1983 & (17)\\
      60 & 60 & $1.276\times 10^{-13}$ (Upper limit) & 1983 & (17)\\
      100 & 100 & $8.022\times 10^{-13}$ (Upper limit) & 1983 & (17)\\
      \hline
      \multicolumn{5}{@{}l@{}}{\hbox to 0pt{\parbox{170mm}{\footnotesize
        Notes.
        \par\noindent
        \footnotemark[$*$] Reference numbers are described in table \ref{tab:PS}.
      }\hss}}
    \end{tabular}
  \end{center}
\end{table}

\begin{table}
  \caption{Means of point shapes in figures \ref{fig:SED1}, \ref{fig:SED2}, and \ref{fig:SED3}.}\label{tab:PS}
  \begin{center}
    \scalebox{0.85}
      {
      \begin{tabular*}{188mm}{llc}
        \hline
        \multicolumn{1}{c}{Point shapes} & \multicolumn{1}{c}{Reference} & \multicolumn{1}{c}{Number}\\
        \hline
        Plus & UCAC4 Catalog \citep{Z} & (1)\\
        Cross & TASS Mark IV patches photometric catalog, version 2 \citep{Dr} & (2)\\
        Star & \citet{Kun} & (3)\\
        Open Box & Catalog of Stellar Spectral Classifications \citep{Sk2013} & (4)\\
        Filled box & Carlsberg Meridian Catalog 14 (CMC14) \citep{Cop} & (5)\\
        Open circle & General Catalog of Variable Stars \citep{Sa} & (6)\\
        Filled circle & Red variables in the NSVS \citep{Wo_b} & (7)\\
        Open triangle & 2MASS All-Sky Catalog of Point Sources \citep{Cu2003} & (8)\\
        Filled triangle & ISLE (our observation) & (9)\\
        Inverted triangle & \citet{Rei} & (10)\\
        Filled inverted triangle & WISE All-Sky Data Release \citep{Cu2012} & (11)\\
        \multirow{2}{*}{Diamond} & C2D Fall'07 High Reliability (HREL) CORES Catalog & (12)\\
        & IRS Enhanced Products & (13)\\
        Filled diamond & MSX6C Infrared Point Source Catalog \citep{E} & (14)\\
        \multirow{2}{*}{Pentagon} & AKARI/IRC mid-IR all-sky Survey \citep{Is} & (15)\\
        & AKARI/FIS All-Sky Survey Point Source Catalogues \citep{Yam} & (16)\\
        Filled pentagon & IRAS catalogue of Point Sources, Version 2.0 \citep{Hel} & (17)\\
        Barred circle & Submillimeter-Continuum SCUBA detections \citep{Di} & (18)\\
        quarter-filled circle & \citet{Rosv} & (19)\\
        \hline
      \end{tabular*}
    }
  \end{center}
\end{table}

\subsection{Observation}
Little is known about the 4 sources selected in the previous section. For example, we looked into the SIMBAD database and no paper was found for 2 sources (IRAS~19574+4941 and V583~Cas.) To identify what these MIR variable sources are, we carried out JHK$\mathrm{_{s}}$-band photometric observations and JHK-band spectroscopic observations from 2013 August 22 to 28 and on 2014 November 20 with the near-infrared imager and spectrograph ISLE (\cite{Yan2006}, \yearcite{Yan2008}), which is attached to Cassegrain focus of the 1.88 m telescope at Okayama Astrophysical Observatory. The detector of ISLE is a 1024 $\times$ 1024 HgCdTe HAWAII array which covers a \timeform{4'.3} $\times$ \timeform{4'.3} field of view with a pixel scale of 0.254 arcsec/pixel. For IRAS~19574+4941, we carried out optical spectroscopic observation at 2014 August 28 with Medium And Low-dispersion Long-slit Spectrograph (MALLS), which is attached to Nasmyth focus of the NAYUTA 2.0 m telescope at Nishi-Harima Astronomical Observatory.

We took defocused images for two of our targets (IRAS~19574+4941 and V583~Cas) because these stars were too bright. Details of imaging observations for individual stars are shown in table \ref{tab:OJI}.

In observation with ISLE, we obtained both low dispersion (R $\sim$ 510, 350, 410 for J, H, K, respectively) and medium dispersion (R $\sim$ 2400, 3600, 2000 for J, H, K, respectively) spectra with a slit of 1.0 arcsec (= 4 pixels) width. We also took Ar and Xe comparison lamps and dome flats. For medium dispersion mode, we obtained dome flats immediately before or after each target exposure to reduce the influence of fringe patterns. We observed A0 stars to calibrate atmospheric transmissions and instrumental efficiency. In observation with MALLS, we obtained low dispersion (R $\sim$ 600) spectra with a slit of 1.2 arcsec width. We also took dark frames, Fe-Ne-Ar comparison lamps, dome flats and a spectrophotometric standard star. Details of spectroscopic observations of individual stars are shown in table \ref{tab:S}. Roughly estimated S/N are also listed in the table.

\begin{table}
  \caption{Summary of imaging observations.}\label{tab:OJI}
  \begin{center}
    \begin{tabular}{llcc}
      \hline
      \multicolumn{1}{c}{Target} & \multicolumn{1}{c}{Date} & \multicolumn{1}{c}{Band} & \multicolumn{1}{c}{Exposure time }\\
       & & & \multicolumn{1}{c}{(second $\times$ number)}\\
      \hline
      & & J & 4 $\times$ 8 \\
      IRAS~19574+4941 & 2013 Aug 26 & H & 4 $\times$ 8 \\
      & & K$\mathrm{_{s}}$ & 4 $\times$ 8 \\
      \hline
      & & J & 4 $\times$ 8 \\
      V2494~Cyg & 2013 Aug 26 & H & 4 $\times$ 8 \\
      & & K$\mathrm{_{s}}$ & 4 $\times$ 8 \\
      \hline
      & & J & 30 $\times$ 8 \\
      2MASS~J22352345+7517076 & 2013 Aug 26 & H & 4 $\times$ 8 \\
      & & K$\mathrm{_{s}}$ & 4 $\times$ 8 \\
      \hline
      & & J & 30 $\times$ 8 \\
      2MASS~J22352442+7517037 & 2013 Aug 26 & H & 4 $\times$ 8 \\
      & & K$\mathrm{_{s}}$ & 4 $\times$ 8 \\
      \hline
      & & J & 30 $\times$ 8 \\
      \textrm{[}RD95\textrm{]}~C & 2013 Aug 26 & H & 4 $\times$ 8 \\
      & & K$\mathrm{_{s}}$ & 4 $\times$ 8 \\
      \hline
      & & J & 30 $\times$ 8 \\
      2MASS~J22352497+7517113 & 2013 Aug 26 & H & 4 $\times$ 8 \\
      & & K$\mathrm{_{s}}$ & 4 $\times$ 8 \\
      \hline
      & & J & 4 $\times$ 8 \\
      V583~Cas & 2013 Aug 26 & H & 4 $\times$ 8 \\
      & & K$\mathrm{_{s}}$ & 4 $\times$ 8 \\
      \hline
    \end{tabular}
  \end{center}
\end{table}

\begin{table}
  \caption{Summary of spectroscopic observations.}\label{tab:S}
  \begin{center}
    \scalebox{0.89}
    {
      \begin{tabular*}{191mm}{llccccc}
        \hline
        \multicolumn{1}{c}{Target} & \multicolumn{1}{c}{Date} & \multicolumn{1}{c}{Dispersion} & \multicolumn{1}{c}{Band} & \multicolumn{1}{c}{Exposure time} & \multicolumn{1}{c}{Standard star} & \multicolumn{1}{c}{minimum S/N\footnotemark[$*$]}\\
        & & & & (second $\times$ number) & &\\
        \hline
        & 2014 Aug 28 & low & optical & 300 $\times$ 9 & V2011 Cyg & 29\\ 
        & & \multirow{2}{*}{low} & J & 20 $\times$ 4 & \multirow{2}{*}{HR 7782} & 69\\
        & & & HK & 20 $\times$ 4 & & 112\\
        IRAS~19574+4941 & 2013 Aug 22 & & J & 30 $\times$ 6 & & 141\\
        & & medium & H & 30 $\times$ 4 & HR 7734 & 77\\
        & & & K & 30 $\times$ 4 & & 132\\
        \hline
        \multirow{2}{*}{V2494~Cyg} & \multirow{2}{*}{2013 Aug 26} & \multirow{2}{*}{medium} & H & 60 $\times$ 20 & \multirow{2}{*}{HR 8246} & 46\\
        & & & K & 60 $\times$ 10 & & 118\\
        \hline
        \multirow{2}{*}{2MASS~J22352345+7517076} & 2013 Aug 25 & low & HK & 30 $\times$ 10 & HR 8844 & 58\\
        & 2013 Aug 28 & medium & K & 120 $\times$ 4 & HR 8598 & 161\\
        \hline
        \multirow{2}{*}{2MASS~J22352442+7517037} & 2013 Aug 25 & low & HK & 120 $\times$ 8 & \multirow{2}{*}{HR 8844} & 106\\
        & 2013 Aug 28 & medium & K & 120 $\times$ 20 & & 118\\
        \hline
        \multirow{2}{*}{2MASS~J22352497+7517113} & 2013 Aug 25 & low & HK & 60 $\times$ 10 & HR 8844 & 151\\
        & 2013 Aug 28 & medium & K & 120 $\times$ 6 & HR 8598 & 138\\
        \hline
        & & \multirow{2}{*}{low} & J & 120 $\times$ 4 & & 29\\
        & & & HK & 30 $\times$ 4 & & 100\\
        \multirow{2}{*}{V583~Cas} & 2013 Aug 22 & & J & 120 $\times$ 10 & \multirow{2}{*}{HR 9019} & 28\\
        & &  medium & H & 60 $\times$ 12 & & 42\\
        & & & K & 30 $\times$ 4 & & 79\\
        & 2014 Nov 20 & low & HK & 60 $\times$ 4 & & 134\\
        \hline
        \multicolumn{7}{@{}l@{}}{\hbox to 0pt{\parbox{183mm}{\footnotesize
          Notes.
          \par\noindent
          \footnotemark[$*$] Without the regions referred in captions for figures \ref{fig:LDSJ}-\ref{fig:MDSK}.
        }\hss}}
      \end{tabular*}
    }
  \end{center}
\end{table}

\subsection{Data reduction}

\subsubsection{Photometry}

Obtained images were reduced with the standard method, i.e. dark frame subtraction (including bias-image subtraction), sky subtraction and flat fielding correction. Then we combined 8 dithered images by taking average with 3$\sigma$ clipping and performed photometry for these combined images. We performed aperture photometry for three stars (IRAS~19574+4941, V2494~Cyg, and V583~Cas) and point spread function (PSF) fitting photometry for the others (2MASS~J22352345+7517076, 2MASS~J22352442+7517037, [RD95]~C, and 2MASS~J22352497+7517113) because they were too crowded to perform aperture photometry.

To convert instrumental magnitudes to calibrated apparent magnitudes, we referenced the 2MASS All-Sky Catalog of Point Sources \citep{Cu2003}. We calculated differences between 2MASS magnitude and instrumental magnitude of all sources detected in the field of view (except the targets) as follows. 
\begin{enumerate}
  \item We extract instrumental fluxes for a target star and in-field calibration sources (hereafter calibrators) in the combined image.
  \item We compute target star's magnitude for each calibrator using
    \begin{equation}
      m_{\mathrm{target}}-m_{\mathrm{cal}}=-2.5\log_{10}\left(\frac{flux_{\mathrm{inst,\: target}}}{flux_{\mathrm{inst,\: cal}}}\right),
    \end{equation}
        where $m_{\mathrm{target}}$ is the magnitude of the target star and $m_{\mathrm{cal}}$ is the magnitude of the chosen calibration source from 2MASS. The $flux_{\mathrm{inst,\: target}}$ is the aperture or PSF-fit flux extracted for the target star while the $flux_{\mathrm{inst,\: cal}}$ is the same for the chosen calibration source.
  \item We compute associated uncertainty for each target star magnitude as follows.
    \begin{equation}
        \sigma_{m\mathrm{,\: target}}\!=\!\sqrt{\sigma_{m\mathrm{,\: cal}}^{2}\!+\!\left(\frac{2.5}{\ln 10}\!\times\!\frac{\sigma_{{flux}_{\mathrm{inst,\: target}}}}{flux_{\mathrm{inst,\: target}}}\right)^{2}\!+\!\left(\frac{2.5}{\ln 10}\!\times\frac{\sigma_{{flux}_{\mathrm{inst,\: cal}}}}{flux_{\mathrm{inst,\: cal}}}\right)^{2}},
    \end{equation}
        where $\sigma_{m\mathrm{,\: target}}$ is the uncertainty on the magnitude of the target star and $\sigma_{m\mathrm{,\: cal}}$ comes from 2MASS for each calibrators. Both $\sigma_{{flux}_{\mathrm{inst,\: target}}}$ and $\sigma_{{flux}_{\mathrm{inst,\: cal}}}$ consist of two error terms. One term is the random error computed by following formula.
    \begin{equation}
      random\ error = \frac{2.5}{\ln{10}} \times \frac{\frac{flux}{gain}+area\times stdev^{2}+\frac{area^{2}\times stdev^{2}}{nsky}}{flux},
    \end{equation}
        where $area$ means the number of pixels in the aperture, $stdev$ means the standard deviation of the sky value, and $nsky$ means the number of pixels used to estimate the sky level (the readout noise is negligible). Our targets are very bright, so Poisson noises are dominant in random errors and other noises make little contribution. The other term is the error on the repeatability of the photometric measurement due to the atmosphere and the instrument. We considered the standard deviation of the $flux_{\mathrm{inst}}$ measurements for the target star and all calibrators as this error term. Some calibrators are too faint compared to the target star and we cannot perform photometry for the calibrators in each individual image. In that case, we substituted $\sqrt{8}\times random\ error$ for the error, where 8 is the number of images combined. We summed the two terms in quadrature.
  \item We average all target star magnitudes, then average all errors. Three iterations of 3$\sigma$ clipping were used to remove outliers. The number of 2MASS sources in each individual image and the number of calibrators are listed in table \ref{tab:DP}.
\end{enumerate}

\begin{table}
  \caption{The number of 2MASS sources seen in each image and 2MASS sources used for photometric calibration.}\label{tab:DP}
  \begin{center}
    \begin{tabular}{lccc}
      \hline
      \multicolumn{1}{c}{Target} & \multicolumn{1}{c}{Band} & \multicolumn{1}{c}{All 2MASS sources} & \multicolumn{1}{c}{2MASS sources used for photometric calibration}\\
      \hline
      & J & 7 & 7\\
      IRAS~19574+4941 & H & 3 & 3\\
      & K$\mathrm{_{s}}$ & 3 & 2\\
      \hline
      & J & 7 & 7\\
      V2494~Cyg & H & 9 & 9\\
      & K$\mathrm{_{s}}$ & 10 & 10\\
      \hline
      & J & 12 & 10\\
      2MASS~J22352345+7517076 & H & 13 & 10\\
      & K$\mathrm{_{s}}$ & 14 & 10\\
      \hline
      & J & 12 & 10\\
      2MASS~J22352442+7517037 & H & 13 & 10\\
      & K$\mathrm{_{s}}$ & 14 & 10\\
      \hline
      & J & 12 & 10\\
      \textrm{[}RD95\textrm{]}~C & H & 13 & 10\\
      & K$\mathrm{_{s}}$ & 14 & 10\\
      \hline
      & J & 12 & 10\\
      2MASS~J22352497+7517113 & H & 13 & 10\\
      & K$\mathrm{_{s}}$ & 14 & 10\\
      \hline
      & J & 8 & 8\\
      V583~Cas & H & 11 & 11\\
      & K$\mathrm{_{s}}$ & 8 & 8\\
      \hline
    \end{tabular}
  \end{center}
\end{table}

\subsubsection{Spectroscopy}

We performed the standard data reduction for all obtained spectra, including sky subtraction (which contained bias and dark frame subtraction), flat fielding, and wavelength calibration for NIR spectra. Flat fielding also has a role to de-fringe for medium dispersion spectra. Although we performed flat-field correction, fringe features are seen in H-band spectrum of V583 Cas. Then we tried Fourier technique to de-fringe, but unfortunately, it did not work well. To correct the atmospheric transmissions and the instrumental efficiency, spectra of our targets were divided by spectra of A0-type stars (HR~7734, HR~7782, HR~8246, HR~8598, HR~8844, and~HR 9019) that were observed at similar airmass. These A0-type stars were observed immediately before or after each target exposure. When we performed this correction, the hydrogen absorption features (Paschen series lines and Brackett series lines) of each star were removed by spectral fitting with Gaussian profile. We assumed that all A0-type stars have spectra of 10000~K black body. For the optical spectrum, we also performed the standard data reduction, which includes dark frame subtraction, flat fielding, sky subtraction, and wavelength calibration. We used a spectrophotometric standard star (V2011 Cyg) to correct the atmospheric transmissions and the instrumental efficiency.

The dispersion of obtained spectra is 5.9 \AA/pixel, 13.4 \AA/pixel, 13.4 \AA/pixel for J, H, K, respectively (low dispersion mode), 1.66 \AA/pixel, 1.63 \AA/pixel, 3.36 \AA/pixel for J, H, K, respectively (medium dispersion mode), and 2.89 \AA/pixel for optical data. The accuracy of wavelength determination in wavelength calibration is better than 1.3 \AA\ in low dispersion mode of ISLE, 0.18 \AA\ in medium dispersion mode of ISLE, and 0.2 \AA\ for MALLS's data.

\section{Results and Discussions}

\begin{figure}
 \begin{center}
  \FigureFile(160mm,192mm){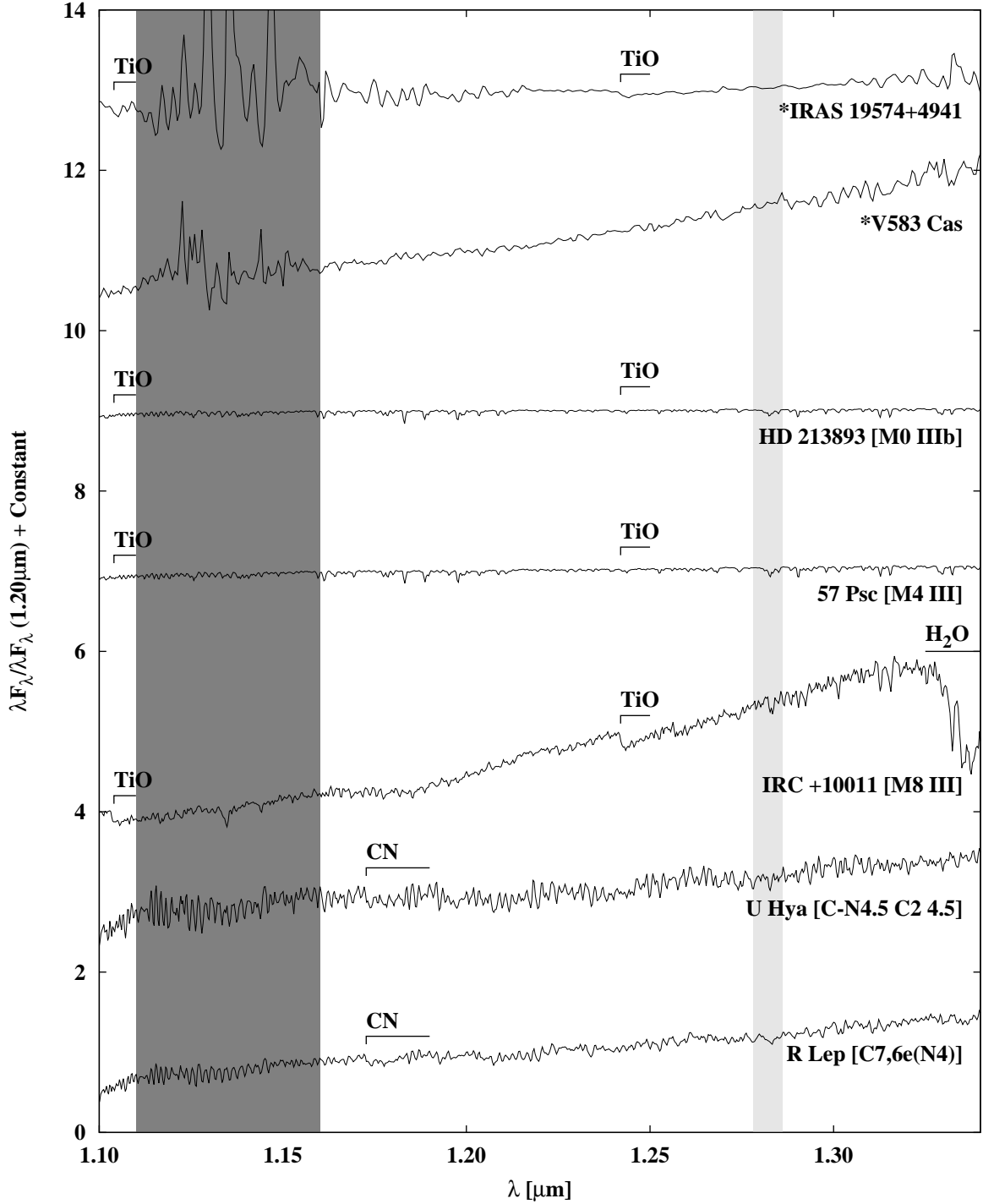}
 \end{center}
 \caption{Low dispersion spectra of our targets in the J-band. For comparison purpose, M-type and carbon stars are also shown from the IRTF Spectral Library \citep{Ra}. The region of strong atmospheric absorption is shown in dark gray. The light gray hatch represents the region where we may not sufficiently correct hydrogen absorption features in the calibration process. Stars with asterisks are our target stars.}
 \label{fig:LDSJ}
\end{figure}

\begin{figure}
 \begin{center}
  \FigureFile(160mm,192mm){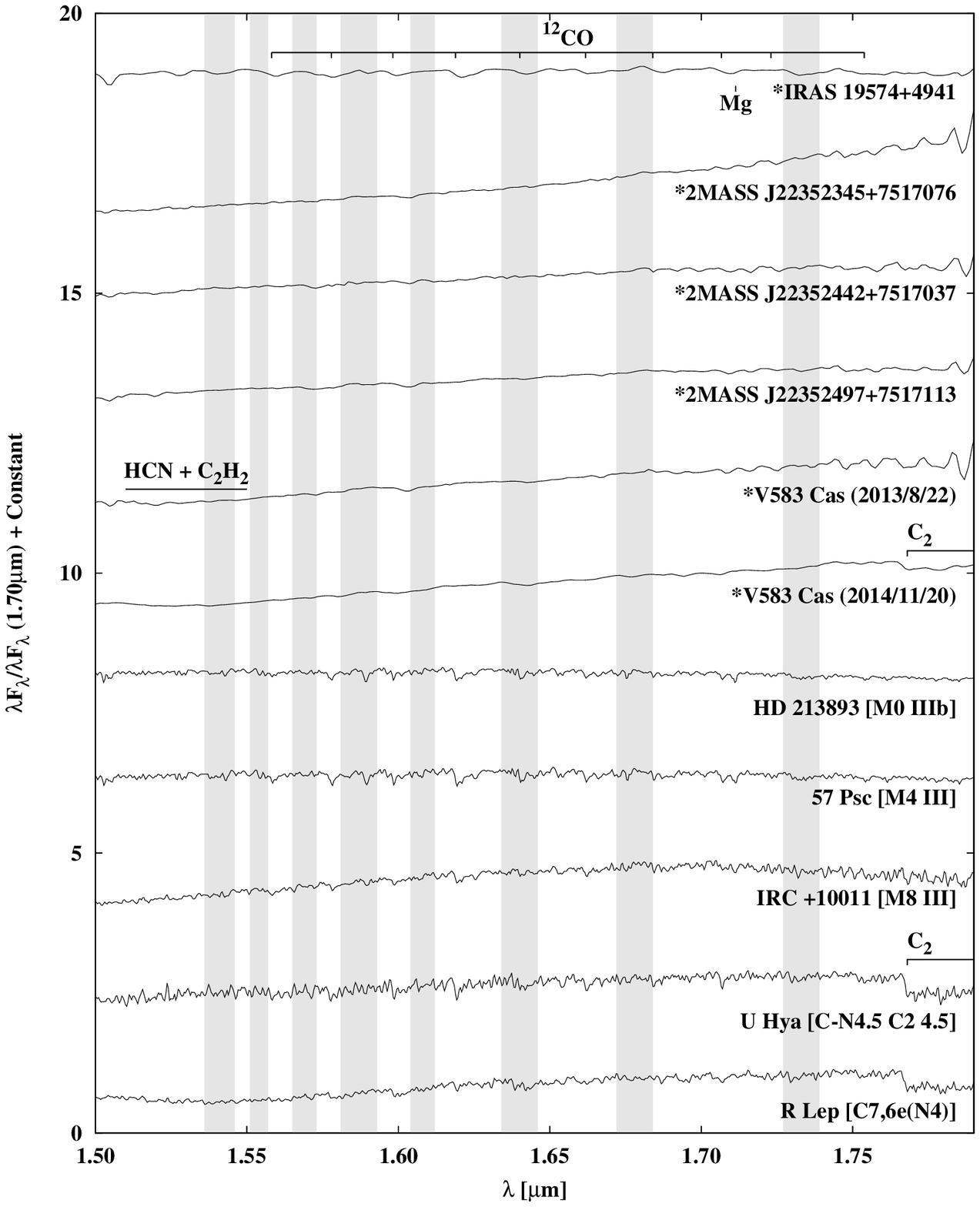}
 \end{center}
 \caption{Low dispersion spectra of our targets in the H-band. For comparison purpose, M-type and carbon stars are also shown from the IRTF Spectral Library \citep{Ra}. The light gray hatch represents the region where we may not sufficiently correct hydrogen absorption features in the calibration process. Stars with asterisks are our target stars.}
 \label{fig:LDSH}
\end{figure}

\begin{figure}
 \begin{center}
  \FigureFile(160mm,192mm){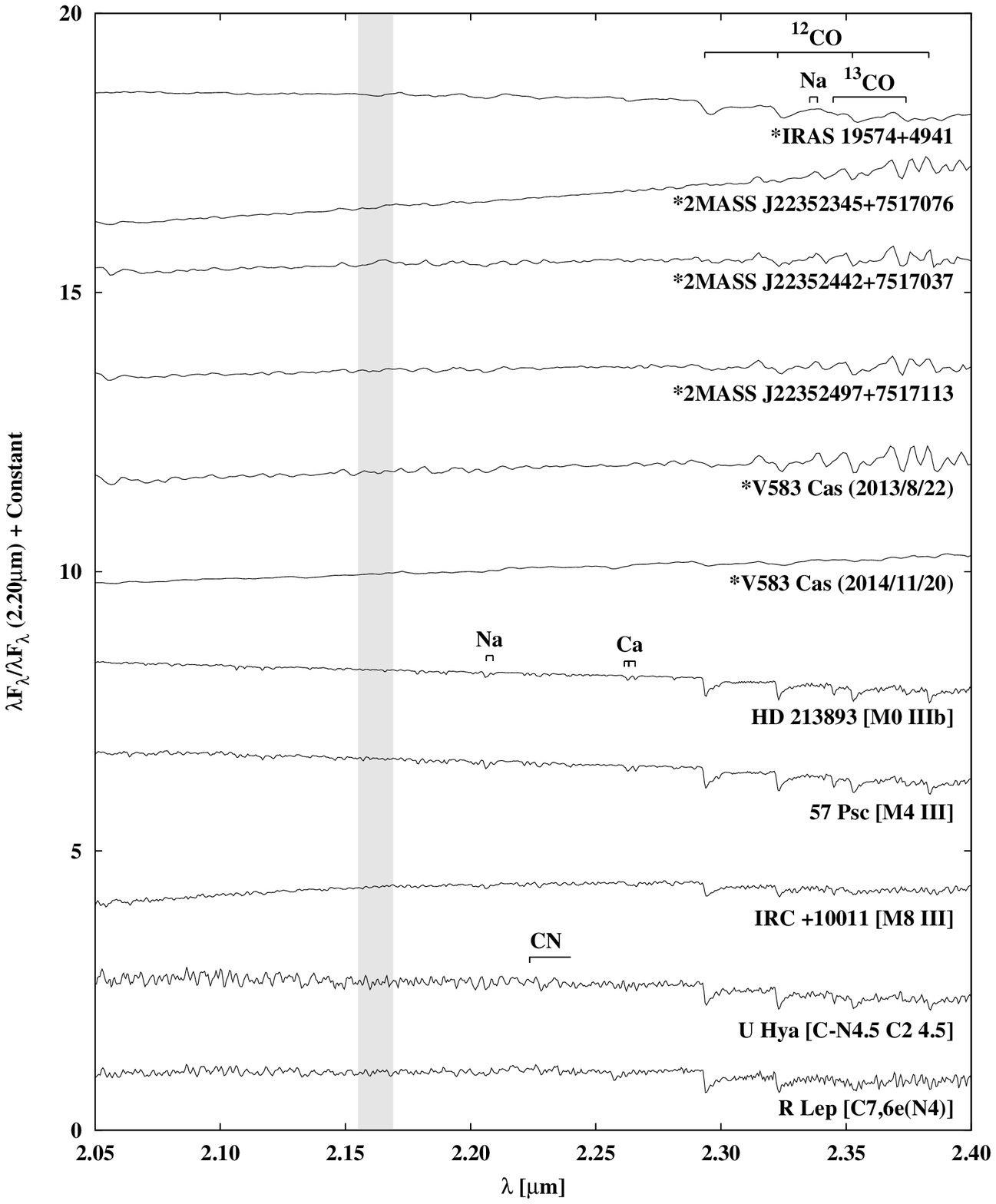}
 \end{center}
 \caption{Low dispersion spectra of our targets in the K-band. For comparison purpose, M-type and carbon stars are also shown from the IRTF Spectral Library \citep{Ra}. The light gray hatch represents the region where we may not sufficiently correct hydrogen absorption features in the calibration process. Stars with asterisks are our target stars.}
 \label{fig:LDSK}
\end{figure}

\begin{figure}
 \begin{center}
  \FigureFile(160mm,192mm){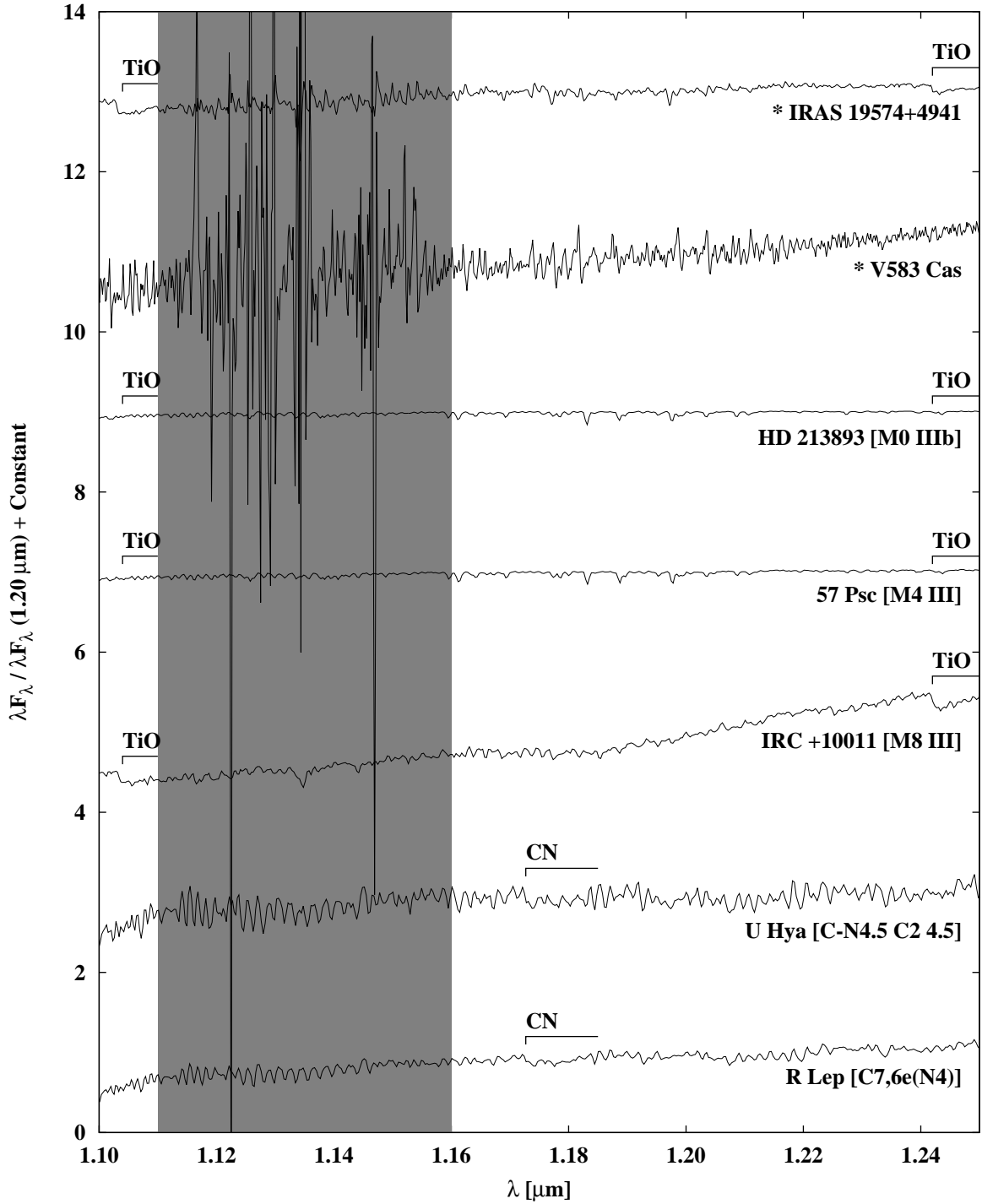}
 \end{center}
 \caption{Medium dispersion spectra of our targets in the J-band. For comparison purpose, M-type and carbon stars are also shown from the IRTF Spectral Library \citep{Ra}. The region of strong atmospheric absorption is shown in dark gray. Stars with asterisks are our target stars.}
 \label{fig:MDSJ}
\end{figure}

\begin{figure}
 \begin{center}
  \FigureFile(160mm,192mm){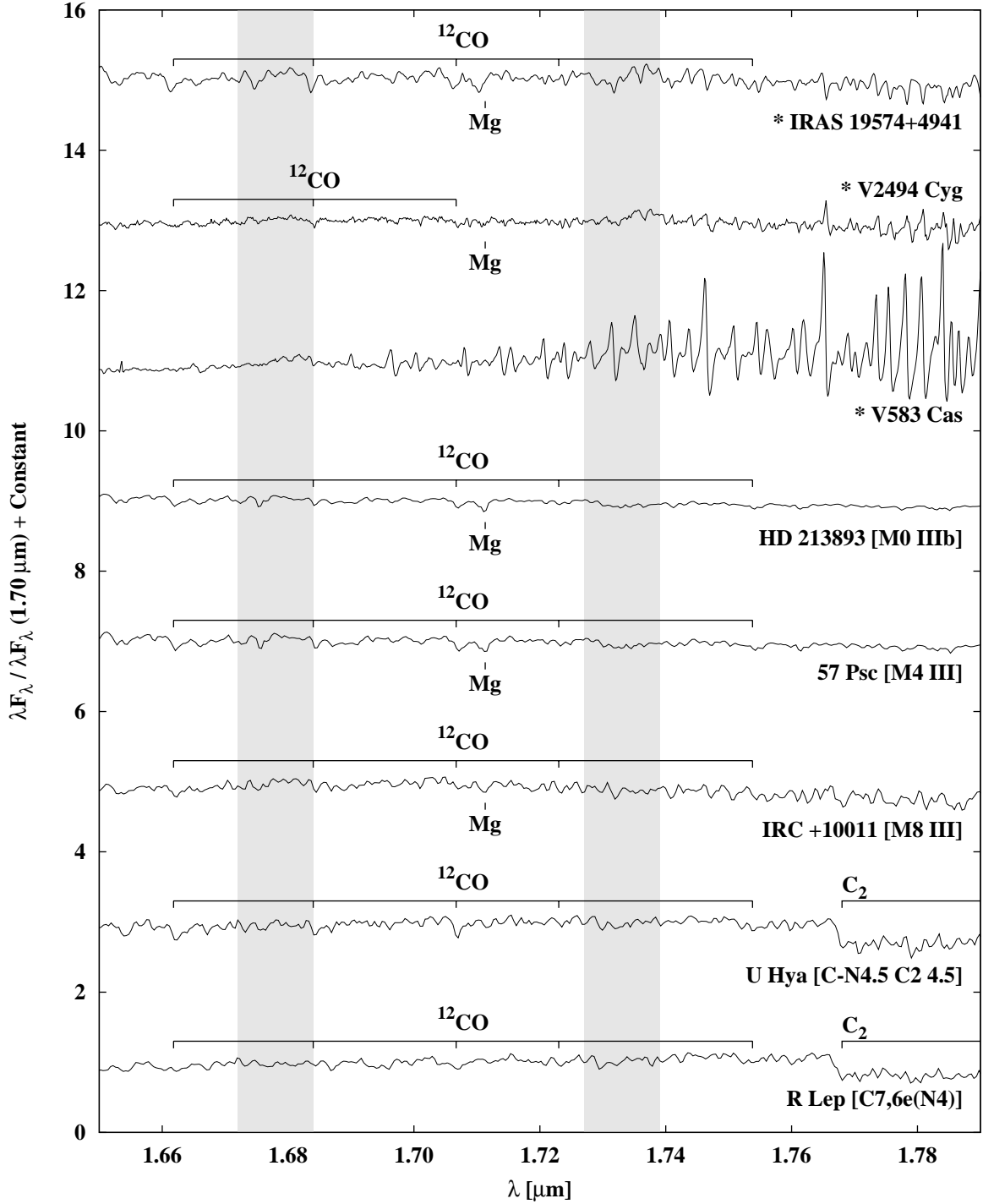}
 \end{center}
 \caption{Medium dispersion spectra of our targets in the H-band. For comparison purpose, M-type and carbon stars are also shown from the IRTF Spectral Library \citep{Ra}. The light gray hatch represents the region where we may not sufficiently correct hydrogen absorption features in the calibration process. Features seen in V583~Cas spectrum longer than 1.7 $\micron$ may be fringe features that we cannot sufficiently correct. Stars with asterisks are our target stars.}
 \label{fig:MDSH}
\end{figure}

\begin{figure}
 \begin{center}
  \FigureFile(160mm,192mm){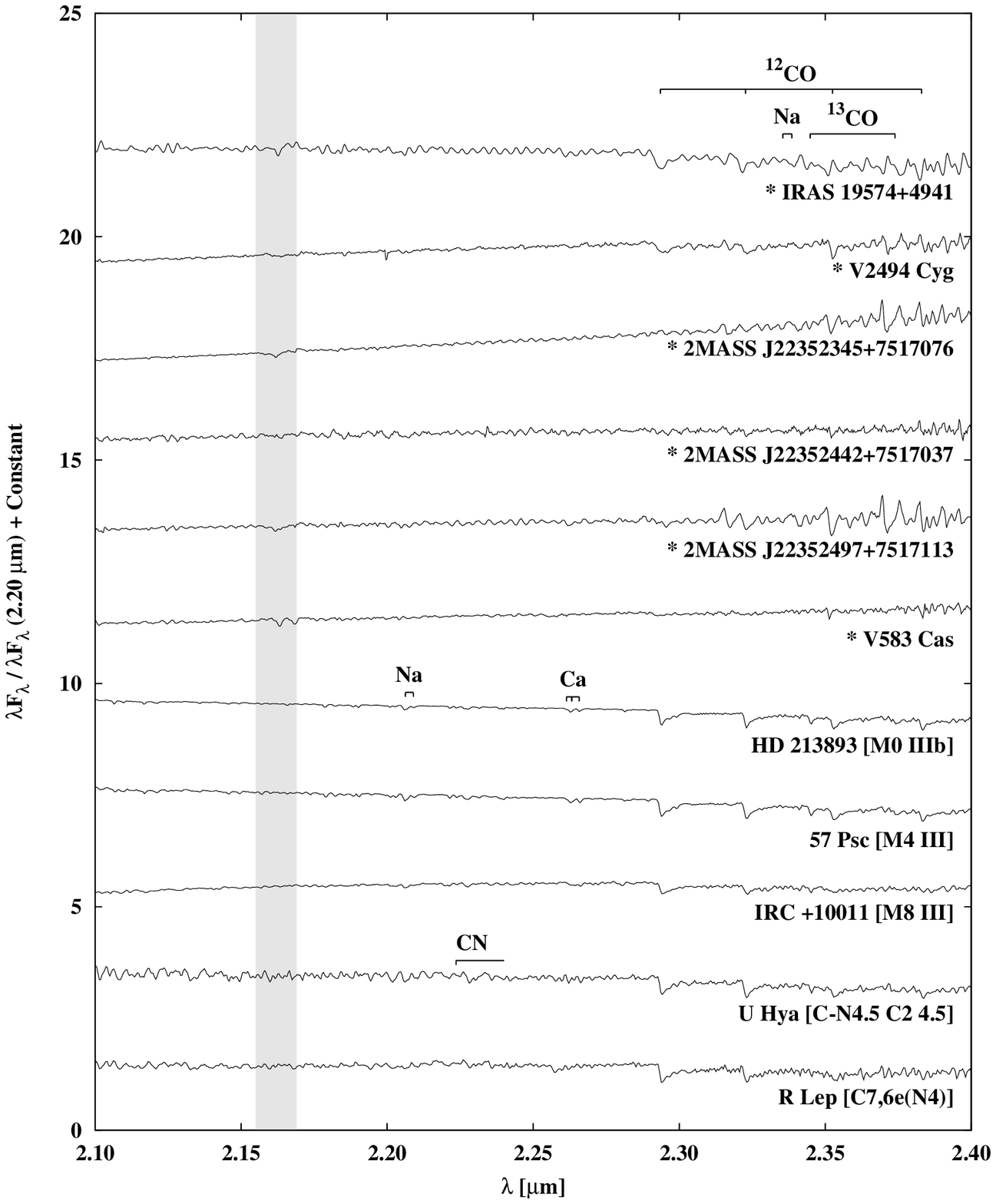}
 \end{center}
 \caption{Medium dispersion spectra of our targets in the K-band. For comparison purpose, M-type and carbon stars are also shown from the IRTF Spectral Library \citep{Ra}. The light gray hatch represents the region where we may not sufficiently correct hydrogen absorption features in the calibration process. Stars with asterisks are our target stars.}
 \label{fig:MDSK}
\end{figure}

\begin{figure}
 \begin{center}
  \FigureFile(160mm,192mm){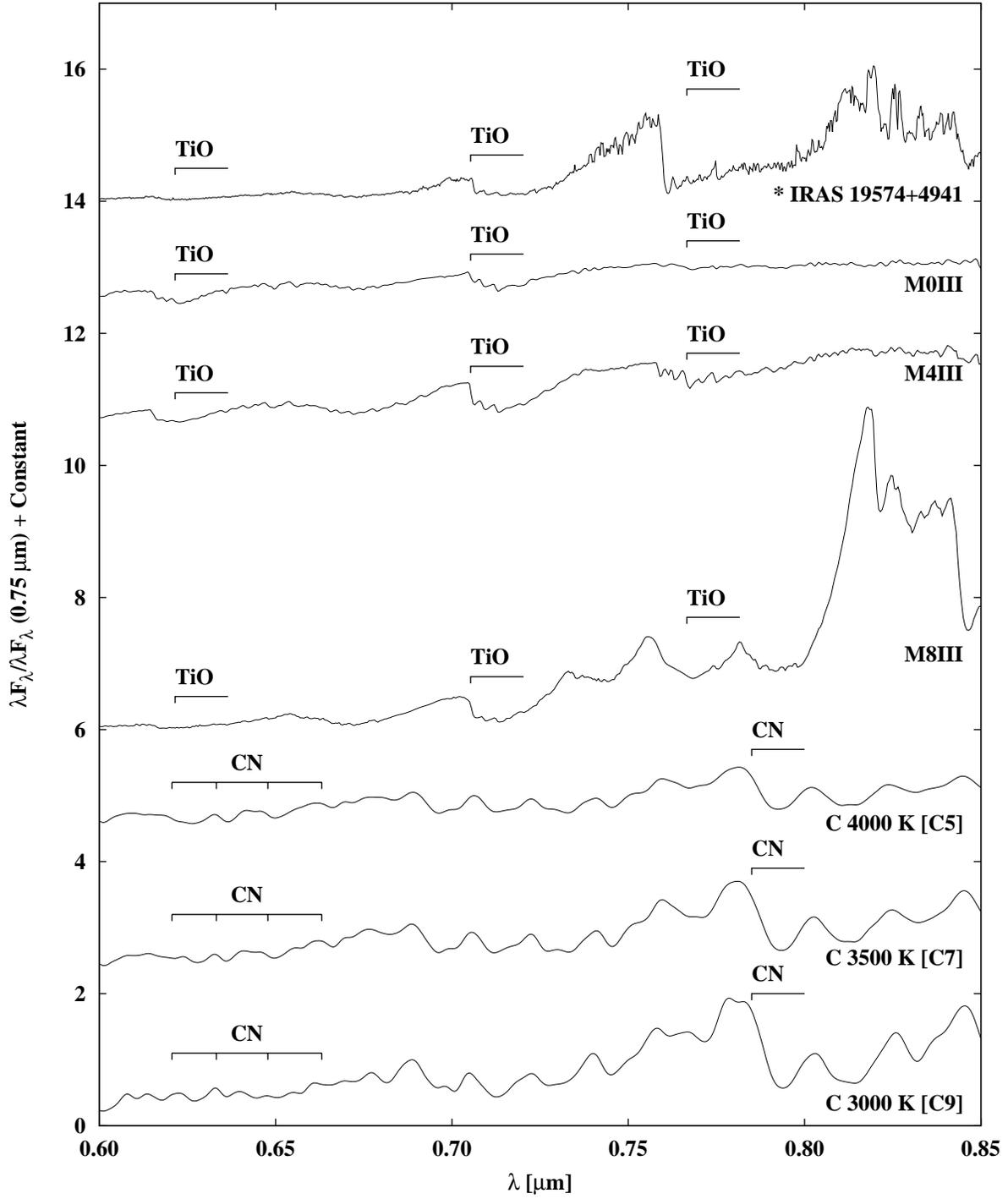}
 \end{center}
 \caption{Low dispersion optical spectra of our targets. For comparison purpose, M-type stars are shown from \citet{Pi} and solar metallicity model spectra of carbon stars are shown from \citet{Ar}. A Star with an asterisk is our target star.}
 \label{fig:LDSO}
\end{figure}

The results of photometric calibration are listed in table \ref{tab:PD}. The reduced spectra of our targets are shown in figures \ref{fig:LDSJ}-\ref{fig:LDSO} and features seen in the spectra are listed in tables \ref{tab:SF1} and \ref{tab:SF2}. We discuss each star below.

\begin{table}
  \caption{Photometric data of our targets.}\label{tab:PD}
  \begin{center}
    \scalebox{0.92}
    {
      \begin{tabular*}{185mm}{lcccccc}
        \hline
        \multicolumn{1}{c}{Target} & \multicolumn{1}{c}{Band} & \multicolumn{1}{c}{ISLE mag} & \multicolumn{1}{c}{ISLE mag uncertainties} & \multicolumn{1}{c}{2MASS mag} & \multicolumn{1}{c}{2MASS uncertainties}\\
        \hline
        & J & 6.442 & 0.091 & 6.397 & 0.026 \\
        IRAS~19574+4941 & H & 5.760 & 0.110 & 5.523 & 0.047 \\
        & K$\mathrm{_{s}}$ & 5.167 & 0.166 & 5.035 & 0.026 \\
        \hline
        & J & 11.586 & 0.127 & 11.544 & 0.031 \\
        V2494~Cyg & H & 9.770 & 0.092 & 9.813 & 0.030 \\
        & K$\mathrm{_{s}}$ & 8.168 & 0.131 & 8.305 & 0.017 \\
        \hline
        & J & 15.056 & 0.224 & 14.230 & Upper limit \\
        2MASS~J22352345+7517076 & H & 10.646 & 0.138 & 12.030 & Upper limit \\
        & K$\mathrm{_{s}}$ & 7.621 & 0.126 & 11.590 & 0.054 \\
        \hline
        & J & 13.612 & 0.107 & 12.957 & Upper limit \\
        2MASS~J22352442+7517037 & H & 11.442 & 0.137 & 11.049 & Upper limit \\
        & K$\mathrm{_{s}}$ & 10.103 & 0.117 & 9.888 & 0.036 \\
        \hline
        & J & 14.365 & 0.106 & &\\
        \textrm{[}RD95\textrm{]}~C & H & 12.781 & 0.132 & & \\
        & K$\mathrm{_{s}}$ & 11.889 & 0.197 & & \\
        \hline
        & J & 11.929 & 0.111 & 11.928 & 0.031\\
        2MASS~J22352497+7517113 & H & 9.677 & 0.142 & 9.714 & 0.031 \\
        & K$\mathrm{_{s}}$ & 8.426 & 0.110 & 8.461 & 0.028\\
        \hline
        & J & 10.634 & 0.044 & 10.527 & 0.023 \\
        V583~Cas & H & 8.320 & 0.051 & 8.338 & 0.023 \\
        & K$\mathrm{_{s}}$ & 6.450 & 0.076 & 6.591 & 0.018 \\
        \hline
      \end{tabular*}
    }
  \end{center}
\end{table}

\begin{table}
  \caption{Spectral features seen in each star.}\label{tab:SF1}
  \begin{center}
    \begin{tabular}{lcc}
      \hline
      \multicolumn{1}{c}{Target} & \multicolumn{1}{c}{Wavelength ($\micron$)} & \multicolumn{1}{c}{Features}\\
      \hline
      & 0.622 & TiO\\
      & 0.705 & TiO\\
      & 0.767 & TiO\\
      & 0.844 & TiO\\
      & 1.104 & TiO\\
      & 1.242 & TiO\\
      & 1.558 & \atom{C}{}{12}O \\
      & 1.578 & \atom{C}{}{12}O \\
      & 1.598 & \atom{C}{}{12}O \\
      & 1.619 & \atom{C}{}{12}O \\
      & 1.640 & \atom{C}{}{12}O \\
      & 1.662 & \atom{C}{}{12}O \\
      IRAS~19574+4941 & 1.684 & \atom{C}{}{12}O\\
      & 1.707 & \atom{C}{}{12}O \\
      & 1.711 & Mg $_{\mathrm{I}}$\\
      & 1.723 & \atom{C}{}{12}O \\
      & 1.754 & \atom{C}{}{12}O \\
      & 2.294 & \atom{C}{}{12}O \\
      & 2.323 & \atom{C}{}{12}O \\
      & 2.335 & Na $_{\mathrm{I}}$\\
      & 2.339 & Na $_{\mathrm{I}}$\\
      & 2.345 & \atom{C}{}{13}O \\
      & 2.353 & \atom{C}{}{12}O \\
      & 2.374 & \atom{C}{}{13}O \\
      & 2.383 & \atom{C}{}{12}O \\
      \hline
      & 1.662 & \atom{C}{}{12}O \\
      & 1.684 & \atom{C}{}{12}O\\
      & 1.707 & \atom{C}{}{12}O \\
      & 1.711 & Mg $_{\mathrm{I}}$\\
      \multirow{2}{*}{V2494~Cyg} & 2.294 & \atom{C}{}{12}O \\
      & 2.323 & \atom{C}{}{12}O \\
      & 2.345 & \atom{C}{}{13}O \\
      & 2.353 & \atom{C}{}{12}O \\
      & 2.374 & \atom{C}{}{13}O \\
      & 2.383 & \atom{C}{}{12}O \\
      \hline
    \end{tabular}
  \end{center}
\end{table}

\begin{table}
  \caption{Continuation of table \ref{tab:SF1}.}\label{tab:SF2}
  \begin{center}
    \begin{tabular}{lcc}
      \hline
      \multicolumn{1}{c}{Target} & \multicolumn{1}{c}{Wavelength ($\micron$)} & \multicolumn{1}{c}{Features}\\
      \hline
      & 2.294 & \atom{C}{}{12}O \\
      & 2.323 & \atom{C}{}{12}O \\
      \multirow{2}{*}{2MASS~J22352442+7517037} & 2.345 & \atom{C}{}{13}O \\
      & 2.353 & \atom{C}{}{12}O \\
      & 2.374 & \atom{C}{}{13}O \\
      & 2.383 & \atom{C}{}{12}O \\
      \hline
      & 2.294 & \atom{C}{}{12}O \\
      & 2.323 & \atom{C}{}{12}O \\
      \multirow{2}{*}{2MASS~J22352497+7517113} & 2.345 & \atom{C}{}{13}O \\
      & 2.353 & \atom{C}{}{12}O \\
      & 2.374 & \atom{C}{}{13}O \\
      & 2.383 & \atom{C}{}{12}O \\
      \hline
      & 1.53 & HCN + C$_{2}$H$_{2}$\\
      & 1.77 & C$_{2}$\\
      \multirow{2}{*}{V583~Cas} & 2.294 & \atom{C}{}{12}O \\
      & 2.323 & \atom{C}{}{12}O \\
      & 2.353 & \atom{C}{}{12}O \\
      & 2.383 & \atom{C}{}{12}O \\
      \hline
    \end{tabular}
  \end{center}
\end{table}

\subsection{IRAS~19574+4941}

This star showed significant but relatively weak MIR light variations among our targets. 

\begin{figure}
 \begin{center}
  \FigureFile(160mm,96mm){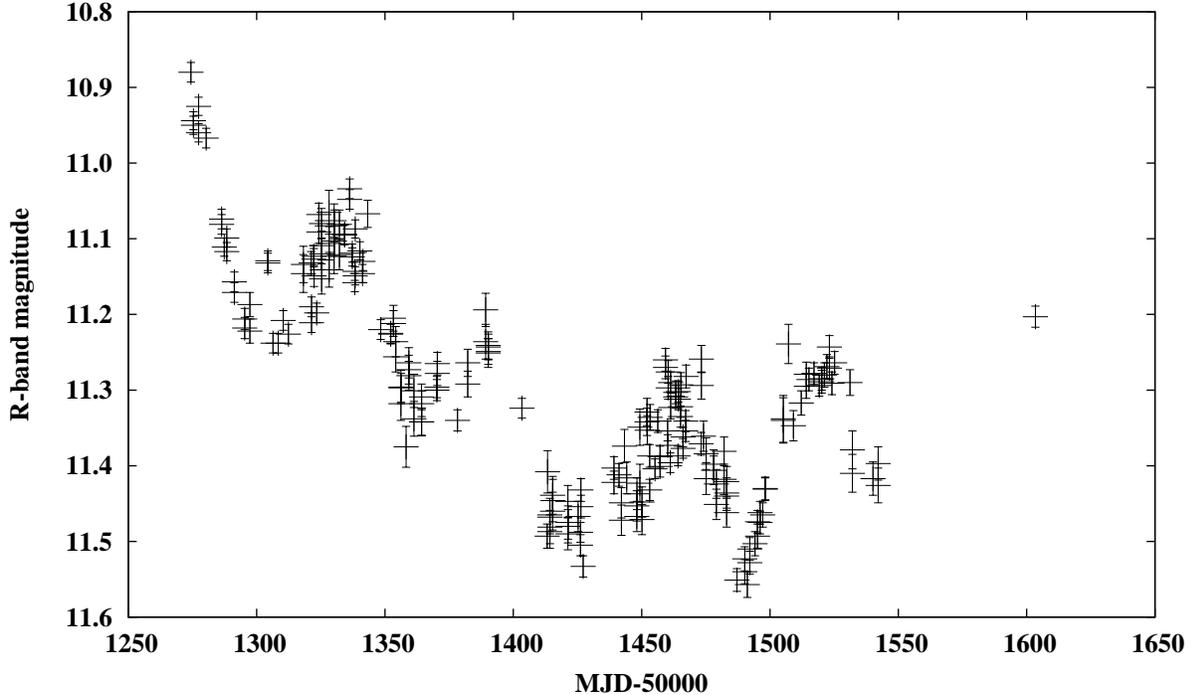}
 \end{center}
 \caption{The light curve of IRAS~19574+4941.}
 \label{fig:LC}
\end{figure}

This star is classified into a semiregular or an irregular variable by \citet{Wo_b} using NSVS \citep{Wo_a} data, and an irregular variable by AAVSO International Variable Star Index VSX \citep{Wat}. The light curve of this star from NSVS catalog is shown in figure \ref{fig:LC}.

This star showed no discernible NIR light variations from 2MASS data. This star showed TiO absorption features in optical and J-band spectra and CO absorption features in H- and K-band. These features are typical of an M-type star's spectrum. Additionally, we calculated reduced chi-squares by comparing this star's spectra with comparison stars' spectra or model spectra (see table \ref{tab:CS}). When we calculated the reduced chi-squares, we interpolated the values of flux densities using cubic spline. The reduced chi-squares were calculated by the following formula,
\begin{equation}
\chi^{2}=\frac{1}{N-1}\sum_{i=1}^{N}\frac{(\lambda_{i} F_{\lambda,\: \mathrm{target},\: i} - \lambda_{i} F_{\lambda,\: \mathrm{comparison\: or\: model},\: i})^{2}}{\lambda_{i} F_{\lambda,\: \mathrm{comparison\: or\: model},\: i}}
\end{equation}
where $N$ is the number of data points, $F_{\lambda,\: \mathrm{target}}$ is the flux density of a target's spectrum, and $F_{\lambda,\: \mathrm{comparison\: or\: model}}$ is the flux density of a comparison star's or model spectrum. The same calculation is done for V583~Cas's spectra. These results suggest that this star is a relatively early M-type star. Therefore, we conclude that this star is an M-type star.

Can this star be a nearby M-dwarf? According to \citet{Cox}, absolute magnitude of the M0V is 5.99, 5.32, and 5.15 for J-, H-, and K-band respectively. By comparing these values with the 2MASS magnitude (table \ref{tab:PD}), we can estimate its distance from the Earth, and it is $\sim$\nobreak 10~pc. However, the proper motion of this star is very small ($\sim$10~mas/yr, \cite{Mon, Roes}), and this contradicts any close distance. Therefore, the star is likely to be more distant and hence could be an evolved star. Additionally, this star is not a member of known star forming regions. Therefore, this star is probably an AGB star. Few monitoring observations for AGB stars in MIR region have been conducted to date. From these observations, the amplitudes of Mira variables in MIR are less than 1~mag \citep{Sm}. However, this star shows stronger brightening than 1~mag at 25~ $\micron$ region, so we cannot explain the reason for brightening just by pulsation. We infer that a large amount of mass loss of the central star is one possible reason for the brightening in MIR. If mass loss rate increases an order of magnitude, we can explain brightening to several times, as we assume symmetric mass loss. The other possibility is that this star can be in a binary system. The light curve of this star in optical is similar to V~Hya \citep{Kn} in terms of having two variability periods. V~Hya is considered to be in a binary system. \citet{Kn} suggested that the shorter period of V~Hya ($\sim$530 days) was due to pulsation and the longer period ($\sim$6000 days) was due to extinction by circumstellar dust orbiting with the companion. Although the periods of these two stars are different (the periods of IRAS~19574+4941 are $\sim$60 days and 500 days), we can consider the possibility that this star is also in a binary system. Also, if this star is in a binary system, we can consider other reasons for brightening in MIR. One possible thing is eclipse, but it has a problem. If the longer period is due to eclipse, the amplitude in R-band ($\sim$0.7 mag) is smaller than the brightening in MIR. Therefore, eclipse would not become main factor of brightening. Another possible thing is mass transfer. When gas and dust accrete to companion, gravitational energy is released in the form of light. As a result, dust around the system absorbs the light and increases the amount of reradiation. Anyway, it needs to confirm whether this star is in a binary system or not to explore the possibility of these things. However, it is difficult to confirm whether a red giant is in a binary system or not because we cannot always distinguish orbital motion and pulsation. If the system has an accretion disk, then ultraviolet spectroscopy should reveal excess continuum or line emission from the accretion shock.

\begin{table}
  \caption{Reduced chi-square values calculated for our targets and comparison stars.}\label{tab:CS}
  \begin{center}
    \scalebox{0.96}
    {
      \begin{tabular*}{177mm}{lccccc}
        \hline
        \multicolumn{1}{c}{Star} & \multicolumn{1}{c}{Dispersion} & \multicolumn{1}{c}{Band} &  \multicolumn{1}{c}{IRAS 19574+4941} & \multicolumn{1}{c}{V583 Cas (2013/8/22)} & \multicolumn{1}{c}{V583 Cas (2014/11/20)}\\
        \hline
        & & optical & 68.63 & & \\
        & \multirow{2}{*}{low} & J & 3.922 & 27.60 & \\
        & & H & 3.333 & 21.10 & 29.00 \\
        M0III & & K & 1.272 & 40.63 & 55.51\\
        & & J & 0.783 & 2.340 & \\
        & medium & H & 1.860 & & \\
        & & K & 5.181 & 16.73 & \\
        \hline
        & & optical & 33.08 & &\\
        & \multirow{2}{*}{low} & J & 3.783 & 25.24 &\\
        & & H & 2.126 & 18.78 & 25.96\\
        M4III & & K & 1.667 & 49.27 & 67.11\\
        & & J & 0.655 & 2.166 & \\
        & medium & H & 1.956 & & \\
        & & K & 4.873 & 21.18 & \\
        \hline
        & & optical & 56.23 & & \\
        & \multirow{2}{*}{low} & J & 42.95 & 12.00 &\\
        & & H & 102.5 & 9.149 & 9.871\\
        M8III & & K & 23.29 & 11.50 & 14.47\\
        & & J & 13.87 & 2.243 &\\
        & medium & H & 17.68 & &\\
        & & K & 15.04 & 3.229 &\\
        \hline
        C5 (4000 K) & low & optical & 82.91 & &\\
        \hline
        C7 (3500 K) & low & optical & 77.31 & &\\
        \hline
        C9 (3000 K) & low & optical & 73.81 & &\\
        \hline
        & & J & 9.050 & 10.62 &\\
        & low & H & 26.80 & 10.27 & 12.11\\
        \multirow{2}{*}{C-N4.5} & & K & 3.817 & 44.06 & 59.07\\
        & & J & 4.234 & 2.828 &\\
        & medium & H & 6.721 & &\\
        & & K & 4.701 & 18.76 &\\
        \hline
        & & J & 8.348 & 10.28 &\\
        & low & H & 39.16 & 4.626 & 5.197\\
        \multirow{2}{*}{C7,6e(N4)} & & K & 6.747 & 24.67 & 32.87\\
        & & J & 2.156 & 1.695 &\\
        & medium & H & 3.267 & &\\
        & & K & 7.755 & 16.15 &\\
        \hline
      \end{tabular*}
    }
  \end{center}
\end{table}

\subsection{V2494~Cyg}

This star showed the strongest brightening in MIR in our samples. Interestingly, strong brightening was also seen in FIR ($F_{\nu,\ \mathrm{AKARI\ N60}} / F_{\nu,\ \mathrm{IRAS\ 60}} = 6.9 \pm 0.5$ and $F_{\nu,\ \mathrm{AKARI\ WIDE-S}} / F_{\nu,\ \mathrm{IRAS\ 100}} = 5.7 \pm 0.8$).

From previous works, this star is considered a FU Orionis star (FUor, e.g. \cite{Rei,Gr}). A FUor is a kind of YSO, which exhibit violent outburst because of a sudden increase of the mass accretion rate (up to $10^{-4} \MO yr^{-1}$) from the accretion disk around it \citep{Ha}. During the outbursts, these stars can increase the bolometric luminosity by two to three orders of magnitude.

This star showed no discernible NIR light variations from 2MASS data. No variation in spectrum of this star from previous data (e.g. \cite{Rei,Gr}) was seen. The deep first overtone CO absorption bands at 2.29 $\micron$ were seen, which is usually observed in the spectrum of FUor type star.

Several photometric monitorings for YSOs in MIR have been conducted to date. \citet{Mor} monitored YSOs in Orion Nebula Cluster and \citet{Fl} surveyed five evolved protoplanetary disks in the IC~348 cluster. They considered the reason for MIR variability as mass accretion, flares, photospheric spots, or variable extinction. The amplitudes of these light variations are smaller than 1 mag and much smaller than that of V2494~Cyg. \citet{Mu} reported multi-epoch MIR observations of protostar LRLL~54361 that exhibits by far the largest MIR flux variability ($\sim$2.5 mag) in IC~348. This protostar showed periodic MIR flux variation caused by pulsed accretion. Large light variation in MIR is also reported by \citet{Fi}. They reported multiwavelength observations of V2775~Ori, a FUor type star, and strong brightening was also caused by increasing mass accretion. To cause such large light variation, it needs to change accretion rate dramatically. Therefore, we think the reason MIR flux was significantly increased is a sudden increase of the mass accretion rate. It caused brightening of the central star and dust around the star was heated rapidly. Heated dust emits more MIR flux. \citet{Ma} suggested that the increase in brightness was started in the early 1980s. Our result indicates that this star started strong brightening between the time when IRAS and AKARI data were taken, restricting the epoch of brightening stronger (after 1983).

Significant light variation of FUors in optical wavelength was well known, but large light variation of MIR and FIR is rarely known. FUors are rarely detected because the brightening timescale of FUors (from a few years to several decades) is much shorter than the timescale between outbursts of the order of $10^{4}$ yr.

\subsection{2MASS~J22352345+7517076, 2MASS~J22352442+7517037, \textrm{[}RD95\textrm{]}~C, and 2MASS~J22352497+7517113}

WISE W3 data showed strong brightening from IRAS 12 data.

These stars are in Cepheus flare region, a star-forming region. \citet{Kun} carried out optical spectroscopic observations for 2MASS~J22352442+7517037 and 2MASS~J22352497+7517113. They classified 2MASS~J22352442+7517037 into an M0-type star, 2MASS~J22352497+7517113 into a K7-type star, and both stars into classical T Tauri stars.

2MASS~J22352345+7517076 showed significant light variations in H and K$\mathrm{_{s}}$-band. Especially, it changed about 4 magnitudes in K$\mathrm{_{s}}$-band but usual AGB stars do not show such large light variation in NIR (Ita \etal , in prep.). It supports the idea that this star is a YSO. The spectrum of this star is nearly featureless and we can barely see CO absorption features in K-band. It is probably due to significant excess emission from a disk of material that veils the stellar features. It also supports that this star is young star with a protoplanetary disk. Therefore, we infer that this star is probably a YSO. 2MASS~J22352442+7517037 and 2MASS~J22352497+7517113 have CO absorption features. It is consistent with \citet{Kun}. 

Dramatic brightening is seen in only WISE data, so this brightening occurred between 2006 and 2009. Because 2MASS~J22352345+7517076 showed significant light variations, this star probably related to the brightening in MIR. However, IRAS, AKARI, and WISE were not able to detect 2MASS~J22352345+7517076, 2MASS~J22352442+7517037, \textrm{[}RD95\textrm{]}~C, and 2MASS~J22352497+7517113 separately. For more detailed discussion, we need the monitoring data that can resolve these stars in MIR.

\subsection{V583~Cas}

This star became $\sim$ 4-5 times brighter in MIR.

This star is classified into LB-type by AAVSO International Variable Star Index VSX and General Catalogue of Variable Stars (GCVS, \cite{Sa}). The spectral type of this star is classified into M-type by \citet{Sk1997}, GCVS, and \citet{Sk2013}. A search through the literature showed that their spectral classification is based on \citet{Rosi}. On the other hand, \citet{It} classified this star into carbon star based on its red color. Both \citet{Kur} and \citet{Al} list this star as a carbon star. However, their data are not publicly available, so it is not possible to confirm the spectral type from previous classifications.

This star also showed no discernible light variations from 2MASS data. In the H-band spectrum of this star, we can see C$_{2}$ Ballik-Ramsay system around 1.77~$\micron$. It is typical of a carbon star's spectrum. In addition to this feature, we can see a weak feature around 1.52 $\micron$. According to \citet{J}, some carbon stars have an absorption band at 1.53 $\micron$ due to HCN and C$_{2}$H$_{2}$. Spectra of these stars show greatly weakened CO absorption features and these stars are very red. These characteristics are consistent with this star ($(J - K_{s})_{\mathrm{2MASS}}=4.215$, $(J - K_{s})_{\mathrm{ISLE}}=3.936$). It also suggests that this star is a carbon star. Hence, we conclude that this star is a carbon star.

We cannot see C$_{2}$ Ballik-Ramsay system in the spectrum taken on 2013 August 22. The reason is probably that absorption by terrestrial water vapor is stronger in summer than in autumn in Japan. The wavelength of the feature is near band end of the H-band filter, so S/N tends to be worse compared to the other wavelengths. Therefore, C$_{2}$ Ballik-Ramsay system could have been obscured by noise, and that is why we could not see the feature.

As described for IRAS~19574+4941, we cannot explain the significant brightening just by pulsation. Therefore, additional mechanisms are needed to explain the MIR brightening. However, observational data for this star are insufficient and we cannot confirm the reason for brightening now.

\section{Summary}

Using IRAS, AKARI, and WISE point source catalogs, we found that 4 sources (IRAS~19574+491, V2494~Cyg, IRAS~22343+7501, and V583~Cas) significantly brightened at MIR wavelengths over the 20-30 years of difference in observing times. IRAS~22343+7501 consists of 4 stars (2MASS~J22352345+7517076, 2MASS~J22352442+7517037, \textrm{[}RD95\textrm{]}~C, and 2MASS~J22352497+7517113) and we obtained first JHK$\mathrm{_{s}}$ photometric data for all 4 sources. Our spectroscopic observation reveals that two stars among the four are evolved stars (one M-type star and one carbon star). The other sources except \textrm{[}RD95\textrm{]}~C (not known the type yet) are young stars (one FUor, two classical T~Tauri star, and one possible YSO). IRAS~19574+4941 and V583~Cas are especially interesting stars because we cannot explain the reason for significant brightening in MIR just by pulsation. We infer that large amounts of mass loss of the central star and/or phenomena related to a binary star system are possible reasons for the brightening in MIR. We need further observations to confirm these inferences.


\begin{thebibliography}{}
\bibitem[Alksnis \etal (2001)]{Al}
  Alksnis,~A., Balklavs,~A., Dzervitis,~U., Eglitis,~I., Paupers,~O., \& Pundure,~I. 2001, BaltA. 10, 1
\bibitem[Aringer \etal (2009)]{Ar}
  Aringer,~B., Girardi,~L., Nowotny,~W., Marigo,~P., \& Lederer,~M.~T. 2009, \aap, 503, 913
\bibitem[Copenhagen University, Obs. \etal (2006)]{Cop}
  Copenhagen~University,~Obs., Institute,~Astronomy~Of, Cambridge, Uk, \& Real Instituto Y Observatorio de La Armada, Fernando En San 2006, VizieR Online Data Catalog, 1304, 0
\bibitem[Cox(2000)]{Cox}
  Cox,~A.~N. 2000, in Allen's Astrophysical Quantities, ed. A.~N.~Cox (New York: Springer), 151
\bibitem[Cutri \etal (2003)]{Cu2003}
  Cutri,~R.~M., \etal\ 2003, VizieR Online Data Catalog, 2246, 0
\bibitem[Cutri \etal (2012)]{Cu2012}
  Cutri,~R.~M., \etal\ 2012, VizieR Online Data Catalog, 2311, 0
\bibitem[Di Francesco \etal (2008)]{Di}
  Di~Francesco,~J., Johnstone,~D., Kirk,~H., MacKenzie,~T., \& Ledwosinska,~E. 2008 \apjs, 175, 277
\bibitem[Droege \etal (2006)]{Dr}
  Droege,~T.~F., Richmond,~M.~W., Sallman,~M.~P., \& Creager,~R.~P. 2006 \pasp, 118, 1666
\bibitem[Duerbeck \& Benetti(1996)]{Du}
  Duerbeck,~H.~W., \& Benetti,~S. 1996, \apj, 468, L111
\bibitem[Egan \etal (2003)]{E}
  Egan,~M.~P., \etal\ 2003, VizieR Online Data Catalog, 5114, 0
\bibitem[Fischer \etal (2012)]{Fi}
  Fischer,~W.~J., \etal\ 2012, \apj, 756, 99
\bibitem[Flaherty \etal (2012)]{Fl}
  Flaherty,~K.~M., Muzerolle,~J., Rieke,~G., Gutermuth,~R., Balog,~Z., Herbst,~W., Megeath,~S.~T., \& Kun,~M. 2012, \apj, 748, 71
\bibitem[Gandhi, Yamamura, and Takita(2012)]{Ga}
  Gandhi,~D., Yamamura,~I., \& Takita,~S. 2012, \apj, 751, L1
\bibitem[Greene \etal (2008)]{Gr}
  Greene,~T.P., Aspin,~C., \& Reipurth,~B. 2008, \aj, 135, 1421
\bibitem[Hartmann \& Kenyon(1985)]{Ha}
  Hartmann,~L., \& Kenyon,~S.~J. 1985, \apj, 299, 462 
\bibitem[Helou \& Walker(1988)]{Hel}
  Helou,~G., \& Walker,~D.~W. ed. 1988, Infrared astronomical satellite (IRAS) catalogs and atlases. Volume 7: The small scale structure catalog (Washington, D.C.: National Aeronautics and Space Administration), 1
\bibitem[Hoard \etal (2009)]{Ho}
  Hoard,~D.~W., Kafka,~S., Wachter,~S., Howell,~S., Brinkworth,~C., Ciardi, D., \& Szkody,~P. 2009, in ASP Conf. Ser. 404, The Eighth Pacific Rim Conference on Stellar Astrophysics: A Tribute to Kam-Ching Leung, ed.\ B.~Soonthornthum, S.~Komonjinda, K.S.~Cheng, \& K.~C.~Leung (San Francisco, CA: ASP), 234
\bibitem[Ishihara \etal (2010)]{Is}
  Ishihara,~D., \etal\ 2010, \aap, 514, A1
\bibitem[Ita \etal (2010)]{It}
  Ita,~Y., \etal\ 2010, \aap, 514, A2
\bibitem[Joyce(1998)]{J}
  Joyce.,~R.~R. 1998, \aj, 115, 2059
\bibitem[Kessler \etal (1996)]{Ke}
  Kessler,~M.~F., \etal\ 1996, \aap 315, L27
\bibitem[Knapp \etal (1999)]{Kn}
  Knapp,~G.~R., Dobrovolsky,~S.~I., Ivezi\'{c},~Z., Young,~K., Crosas,~M., Matter,~J.~A., \& Rupen, M.~P. 1999, \aap, 351, 97
\bibitem[Kun \etal (2009)]{Kun}
  Kun,~M., Balog,~Z., Kenyon,~S.~J., Mamajek,~E.~E., \& Gutermuth,~R.~A. 2009, \apj, 185, 451
\bibitem[Kurtanidze \& Nikolashvili(1989)]{Kur}
  Kurtanidze,~O.~M., \& Nikolashvili,~M.~G. 1989, in Astrophysics, Volume 31, Low dispersion spectral sky survey to find faint carbon stars. IV. Region 90$^{\circ}\ \le$ l $\le$ 115$^{\circ}$, -5$^{\circ}\ \le$ b $\le$ +5$^{\circ}$, ed.\ D.~M.~Sedrakyan (Berlin and Heidelberg: Springer), 714
\bibitem[Magakian \etal (2013)]{Ma}
  Magakian,~T.Y., \etal\ 2013, \mnras, 432, 2685
\bibitem[Meixner \etal (2006)]{Mei}
  Meixner,~M., \etal\ 2006, \aj, 132, 2268
\bibitem[Melis \etal (2012)]{Mel}
  Melis,~C., Zuckerman,~B., Rhee,~J.~H., Song,~I., Murphy,~S.J., \& Bessell,~M.~S. 2012, \nat, 487, 74
\bibitem[Monet \etal (2003)]{Mon}
  Monet,~D.~G., \etal\ 2003, \aj, 125, 984
\bibitem[Morales-Calder\'{o}n \etal (2011)]{Mor}
  Morales-Calder\'{o}n,~C. \etal, 2011, \apj, 733, 50
\bibitem[Muzerolle \etal (2013)]{Mu}
  Muzerolle,~J., Furlan,~E., Flaherty,~K., Balog,~Z., \& Gutermuth,~R. 2013, \nat, 493, 378
\bibitem[Neugebauer \etal (1984)]{N}
  Neugebauer,~G., \etal\ 1984, \apj, 278, 1
\bibitem[Pickles(1998)]{Pi}
  Pickles,~A.~J., 1998, \pasp, 110, 863
\bibitem[Price \etal (2001)]{Pr}
  Price,~S.~D., Egan,~M.P., Carey,~S.J., Mizuno,~D.~R., \& Kuchar,~T.~A. 2001, \aj, 121, 2819
\bibitem[Rayner \etal (2009)]{Ra}
  Rayner,~J.~T., Cushing,~M.~C., \& Vacca,~W.~D. 2009, \apjs, 185, 289
\bibitem[Rebull(2011)]{Reb}
  Rebull,~L.~M. 2011, in ASP Conf. Ser. 448, 16th Cambridge Workshop on Cool Stars, Stellar Systems, and the Sun, ed.\ C.~M.~Johns-Krull, M.~K.~Browning, \& A.~A.~West (San Francisco, CA: ASP), 5
\bibitem[Reipurth \& Aspin(1997)]{Rei}
  Reipurth,~B. \& Aspin,~C. 1997, \aj, 114, 2700
\bibitem[Roeser \etal (2010)]{Roes}
  Roeser,~S., Demleitner,~M., Schilbach,~E. 2010, \aj, 139, 2440
\bibitem[Rosino, Bianchini, \& di Martino(1976)]{Rosi}
  Rosino,~L., Bianchini,~A., and di~Martino,~D. 1976, \aaps, 24, 1
\bibitem[Rosvick \etal (1995)]{Rosv}
  Rosvick,~J.M. \& Davidge,~T.~J. 1995, \pasp, 107, 49
\bibitem[Samus \etal (2009)]{Sa}
  Samus,~N.~N., \etal\ 2009, VizieR Online Data Catalog, 1, 2025
\bibitem[Skiff(1997)]{Sk1997}
  Skiff,~B.~A., 1997, IBVS, 4441, 1
\bibitem[Skiff(2013)]{Sk2013}
  Skiff,~B.~A., 2013, VizieR Online Data Catalog, 1, 2023
\bibitem[Smith \etal (2002)]{Sm}
  Smith,~B.~J., Leisawitz,~D., Castelaz,~M.~W., \& Luttermoser,~D. 2002, \aj, 123, 948
\bibitem[Vijh \etal (2009)]{V}
  Vijh,~U.~P., \etal\ 2009, \aj, 137, 3139
\bibitem[Watson \etal (2013)]{Wat}
  Watson,~C., Henden,~A.~A., \& Price,~A. 2013, VizieR Online Data Catalog, 1, 2027
\bibitem[Wo\'zniak \etal (2004a)]{Wo_a}
  Wo\'zniak, \etal\ 2004, \aj, 127, 2436
\bibitem[Wo\'zniak \etal (2004b)]{Wo_b}
  Wo\'zniak,~P.~R., Williams,~W.~T., Vestrand,~W.~T., \& Gupta,~V. 2004, \aj, 128, 2965
\bibitem[Wright \etal (2010)]{Wr}
  Wright,~E.~L., \etal\ 2010, \aj, 140, 1868
\bibitem[Yamamura \etal (2010)]{Yam}
  Yamamura,~I., Makiuti,~S., Ikeda,~N., Fukuda,~Y., Oyabu,~S., Koga,~T., \& White,~G.J. 2010, VizieR Online Data Catalog, 2298, 0
\bibitem[Yanagisawa \etal (2006)]{Yan2006}
  Yanagisawa,~K., \etal\ 2006, \procspie, 6269, 62693
\bibitem[Yanagisawa \etal (2008)]{Yan2008}
  Yanagisawa,~K., \etal\ 2008, \procspie, 7014, 701437
\bibitem[Zacharias \etal (2012)]{Z}
  Zacharias,~N., Finch,~C.~T., Girard,~T.M., Henden,~A., Barlett,~J.~L., Monet,~D.~G., \& Zacharias,~M.~I. 2012, VizieR Online Data Catalog, 1322, 0
\end{thebibliography}
\end{document}